# The open cluster Pismis 11 and the very luminous blue supergiant HD 80077,⋆

## I. Physical parameters of the cluster

A. Marco[1] and I. Negueruela[1]

Departamento de Física, Ingeniería de Sistemas y Teoría de la Señal. Escuela Politécnica Superior. University of Alicante. Apdo.99 E-03080. Alicante. (Spain)
e-mail: tobarra@dfists.ua.es

**ABSTRACT**

*Context.* The very luminous blue supergiant HD 80077 has been claimed to be a member of the young open cluster Pismis 11, and hence a hypergiant. Membership of the cluster would mean that it is one of the brightest stars in the Galaxy, and one of the few evolved very massive stars whose distance can be accurately determined.
*Aims.* We carry out a comprehensive study of the open cluster Pismis 11, which allows us to derive with accuracy its distance and reddening.
*Methods.* We obtained *UBVRI* photometry of the cluster field and low-resolution spectroscopy of a number of putative members. We derive spectral types from the spectra and determine that the reddening in this direction is standard. We then carry out a careful photometric analysis that allows us to determine individual reddening values, deriving unreddened parameters that are used for the main sequence fit.
*Results.* We identify 43 likely members of Pismis 11 and determine individual reddenings. We study the variation of extinction across the face of the cluster and find some dispersion, with a trend to higher values in the immediate neighbourhood of HD 80077. We estimate a distance of 3.6 kpc for the cluster. If HD 80077 is a member, it has $M_{\rm bol} < -10.5$ and it is one of the three visually brightest stars in the Galaxy. Several early type stars in the vicinity of Pismis 11 fit well the cluster sequence and are likely to represent an extended population at the same distance. About 18′ to the North of Pismis 11, we find a small concentration of stars, which form a clear sequence. We identify this group as a previously uncatalogued open cluster, which we provisionally call Alicante 5. The distance to Alicante 5 is also 3.6 kpc, suggesting that these two clusters and neighbouring early-type stars form a small association.
*Conclusions.* We have identified a small association around Pismis 11, located at a distance of 3.6 kpc. Based on its proper motion, HD 80077 is not a runaway star and may be a member of the cluster. If this is the case, it would be one of the brightest stars in the Galaxy.

**Key words.** open clusters and associations: individual: Pismis 11 – stars: early-type – stars: Hertzsprung-Russell (HR) and C-M diagrams

## 1. Introduction

In spite of their rarity, blue hypergiants (BHGs) are of enormous astrophysical interest. As they are the visually brightest stars, $M_V \leq -9$, they can be seen to huge distances and hence used to derive chemical abundances in distant galaxies. Moreover they can be used as standard candles as far as the Virgo cluster if their luminosity-wind momentum relationship is well calibrated (Kudritzki & Puls 2000) or if the flux-weighted gravity method is extrapolated to their luminosities (Kudritzki et al. 2003).

Unfortunately, our knowledge of the evolutionary status of BHGs is so rudimentary that we are not yet in a position to exploit these capabilities. At present, it is still unclear whether these objects represent extremely massive stars ($M_* \gtrsim 60 M_\odot$) moving away from the main sequence towards the Humphreys-Davidson limit or more moderately massive stars ($M_* \approx 40 M_\odot$) looping back to the blue after a red supergiant phase. If the latter is true, their measured abundances would not reflect at all their initial compositions, preventing their use as tracers of metallicity. Of course, the possibility that both types of objects can be seen as BHGs also exists.

The evolutionary status of BHGs also has a bearing on our understanding of massive star evolution. Massive stars enrich the Galaxy with large amounts of processed material (they are, for example, the main contributors of oxygen) and also inject enormous amounts of energy and momentum into the interstellar medium during their brief life and subsequent death as supernovae. However, our understanding of such processes is only qualitative, as we still lack basic knowledge about mass loss mechanisms in evolved massive stars. Stars with ($M_* \geq 40 M_\odot$) are believed to lose their H-rich mantles as they evolve off the main sequence to become H-depleted Wolf-Rayet stars, but their precise evolutionary path(s) through the diverse population of transitional objects - such as Red Supergiants, Blue and Yellow Hypergiants and Luminous Blue Variables - is uncertain.

There is only a handful of BHGs in the Milky Way (van Genderen 2001). The best studied BHGs are $\zeta^1$ Sco and Cyg OB2 #12, whose luminosities can be known with some accu-racy, as they belong to OB associations. Unfortunately, for many other evolved massive stars, parameters are very difficult to constrain because their distances are not known accurately. One very favourable case is the proposed B2 Ia$^+$ hypergiant HD 80077, which may be associated with the open cluster Pismis 11.

The membership of HD 80077 in Pismis 11 was first proposed by Moffat & Fitzgerald (1977), who also carried out the first photoelectric investigation of the cluster, detecting only seven other members. From them, they derived $d = 3.5$ kpc. If HD 80077 is at the same distance, it has $M_V \approx -9.5$ and $M_{bol} \approx -11$. HD 80077 is only 2′ away from the cluster centre, closer than some proposed members. The reddening across the face of Pismis 11 is variable and was suggested to increase towards HD 80077, further strengthening the likelihood of its membership. However, from an analysis of the H$\beta$ line, Knoechel & Moffat (1982) concluded that the mass loss from HD 80077, though larger than that of a normal B2 Ia supergiant, did not appear exceptionally high. Moreover, the star did not display the typical photometric variability of other blue and yellow hypergiants.

Motivated by this, Carpay et al. (1991) carried out different tests on the luminosity of HD 80077, using UV spectra and optical photometry, and concluded that all estimates, except for this lack of variability, strongly suggested that HD 80077 was extremely luminous. Since their work, the issue has remained unsettled. The parameters of Pismis 11 are also poorly constrained, as the distance estimate is based on photometry of only 7 candidate members, under the assumption of standard reddening.

In this article, we carry out a global study of the open cluster Pismis 11 and the BHG candidate HD 80077. We determine accurate parameters for Pismis 11 and estimate the luminosity for HD 80077. In a future paper, we will use intermediate resolution spectra of HD 80077 and the brightest OB members of Pismis 11 to derive more accurate stellar parameters and abundances.

## 2. Observations and data

### 2.1. The cluster Pismis 11

We obtained $UBVRI$ photometry of Pismis 11 and spectroscopy of its brightest members (as well as some catalogued OB stars in its immediate neighbourhood) using the multi-mode capabilities of ESO Multi-Mode Instrument (EMMI) on the New Technology Telescope (NTT) at the La Silla Observatory (Chile) on the nights of 15 – 17 February 2006.

The instrument was equipped with three thin, back-illuminated and AR-coated CCD cameras: two of them, which are arranged in a mosaic, in the red arm and one in the blue arm. The mosaic in the red arm covers a field of view of 9′.1 × 9′.1 and has a pixel scale of 0′′.166. We used it for the $V$, $R$ and $I$ filter observations. The blue arm has a field of view of 6′.2 × 6′.2 and a pixel scale of 0′′.37 , and was used for the $U$ and $B$ filters.

We obtained low-intermediate resolution spectra of stars in the region using grism #5 in the red arm. This gives a resolving power $R = 1100$ over the 3000–7000Å range. A few bright sources were also observed at higher resolution, using grating #3 in the blue arm ($R = 3400$ over the 3925–4380Å range) and the echelle grating #9 cross-dispersed with grism #3 in the red arm ($R = 10000$ over the range 4000–7900Å with small gap around 4950Å). These higher resolution data will be used in a forthcoming paper.

In addition, we observed HD 80077 with EMMI on June 6th 2003. On this occasion, we used grating #12 in the blue arm

**Table 1.** Log of the photometric observations taken at the NTT on February 2006 for Pismis 11

| Pismis 11 | RA = 09$h$15$m$52.8$s$ | DEC = $-$50°00′43.9′′ |
|---|---|---|
| | (J2000) | (J2000) |

| | Exposure times (s) | | |
|---|---|---|---|
| Filter | Long times | Intermediate times | Short times |
| U | 600 | 200 | 20 |
| B | 90 | 30 | 2 |
| V | 100 | 30 | 2 |
| R | 100 | 30 | 2 |
| I | 100 | 30 | 2 |

(range 4000 – 4900 Å, $R = 1700$) and grating #7 (range 6400 – 7800 Å, $R = 2600$) in the red arm.

Two standard fields from the list of Landolt (1992), SA 98 & SA 101, were observed several times during the night in order to trace extinction and provide standard stars for the transformation. Their images were processed for bias and flat-fielding corrections with the standard procedures using the CCDPROC package in IRAF[1]. Aperture photometry using the PHOT package inside DAOPHOT (IRAF, DAOPHOT) was developed on these fields with different apertures for each filter: 18 pixels for $U$ and $B$ (blue CCD) and 30 pixels for $V$, $R$ and $I$ filters (red mosaic).

Images of Pismis 11 were taken using 3 series of different exposure times to obtain accurate photometry for a magnitude range. They are presented in Table 1.

The reduction of the images of Pismis 11 was done with IRAF routines for the bias and flat-field corrections. Photometry was obtained by point-spread function (PSF) fitting using the DAOPHOT package (Stetson 1987) provided by IRAF. The apertures used are: 15 pixels for $U$, 14 pixels for $B$ and 30 pixels for all the filters observed with the red arm. In order to construct the PSF empirically, we automatically select bright stars (typically 25 stars). After this, we review the candidates and we discard those that do not reach the best conditions for a good PSF star. Once we have the list of PSF stars ($\approx$ 20), we determine an initial PSF by fitting the best function between the 5 options offered by the PSF routine inside DAOPHOT. We allow the PSF to be variable (in order 2) across the frame to take into account the systematic pattern of PSF variability with position on the chip.

We needed to perform aperture correction only for the $U$ and $B$ filters, as the apertures for the $V$, $R$ and $I$ filters were the same (30 pixels) for the standard and programme stars. Finally, we obtained the instrumental magnitudes for all stars. Using the standard stars, we carried out the atmospheric extinction correction and we transformed the instrumental magnitudes to the standard system using the PHOTCAL package inside IRAF.

All spectroscopic data were reduced using the Starlink software packages CCDPACK (Draper et al. 2000) and FIGARO (Shortridge et al. 1997) and analysed using FIGARO and DIPSO (Howarth et al. 1998).

We have completed our dataset with $JHK_s$ photometry from the 2MASS catalogue (Skrutskie et al. 2006).

The number of stars that we could detect in all filters is limited by the long exposure time in the $U$ filter. We show these stars on the $U$-band image in Fig. 1. In Table 2 we list their $X$ and $Y$ positions as seen in Fig. 1, and their identification with ob-jects in the USNO-B1 catalogue, together with their coordinates (right ascension (RA) and declination (DEC) in J2000). A few stars have no obvious identification in the USNO-B1 catalogue. We have identified these objects in the 2MASS catalogue and give their RA and DEC coordinates as assigned in this catalogue. The designation of each star is given by the number indicated on the image (Fig. 1).

---

[1] IRAF is distributed by the National Optical Astronomy Observatories, which are operated by the Association of Universities for Research in Astronomy, Inc., under cooperative agreement with the National Science Foundation

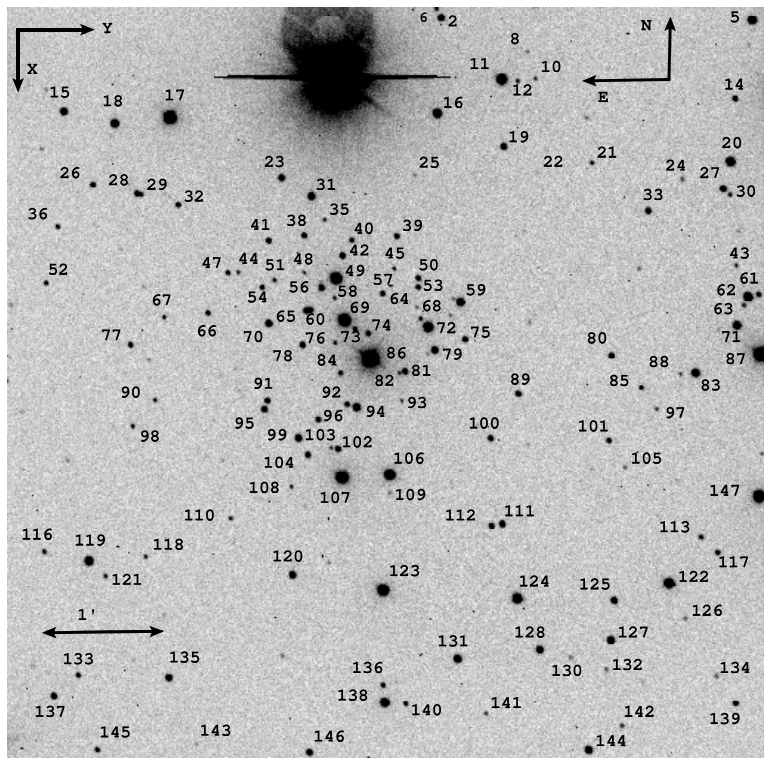

**Fig. 1.** Finding chart for stars in the field of Pismis 11. *XY* positions are listed in Table 2, where they are correlated to RA and DEC. The origin of coordinates is located at the top left corner of the image. The bright star at the top is HD 80077.

We have photometry for 147 stars in the field. In Table 3, we list the values of $V$, $(V-R)$, $(V-I)$, $(U-B)$, and $B$ with the standard deviation and the number of measurements for each magnitude and index. Fig. 1 is the map of the cluster showing the identification of each star for which we have photometry in Table 3.

### 2.2. The new cluster Alicante 5

We obtained $UBVRI$ photometry of a region located near Pismis 11 in order to check if an apparent stellar grouping could be a cluster located at the same distance as Pismis 11. The observations were obtained in service mode on the night of 2007 January 28th with the Superb-Seeing Imager (SuSI2) on the NTT. The size of the field is $5'.5 \times 5'.5$ and the pixel scale is $0''.08$/pixel. The log of the observations is given in Table 4.

The observation procedure for $I$-band observations was different from the rest of the filters. We took five 10-s images with a dithering in the X and Y coordinates. This process is necessary to remove fringing from the images. The final image in the $I$ filter is the sum of these 5 images without fringing. Apart from this, the reduction procedure is identical to that used for the EMMI data, except for the fact that all the filters were observed with the same CCD.

The apertures used are: 50 pixels for all filters for standard stars and 32 pixels in $U$, 34 pixels for $B$, $V$ and $R$ and 30 pixels

**Table 4.** Log of the photometric observations taken at the NTT in January 2007 for Alicante 5

| Alicante 5 | RA = 09h16m27.0s (J2000) | DEC = −49°42′48.9″ (J2000) |
|---|---|---|
| | Filter | Exposure times (s) |
| | U | 600 |
| | B | 90 |
| | V | 60 |
| | R | 60 |
| | I | 50 |

for $I$ in all the target frames. All other steps were done as in the previous section.

After reduction, we have photometry for 105 stars in the field. Table 5 lists values for $V$, $(B-V)$, $(U-B)$, $(V-R)$ and $(V-I)$ with the standard deviation and the number of measures for each magnitude and index. Figure 2 is the map of the cluster showing the identification for all the stars with photometry in Table 6. When mentioned in the text, stars from this frame will be named with an "A" preceding their number to distinguish them from stars in Pismis 11.

In addition, we have spectra for the three brightest stars (A47, A55 and A62 in Table 7) in the grouping, which were taken as part of the EMMI observations.

## 3. Results

### 3.1. The open cluster Pismis 11

#### 3.1.1. Spectroscopy

We have obtained spectra for the brightest stars in Pismis 11. Echelle spectroscopy was taken for the three brightest members given by Moffat & Fitzgerald (1977): HD 80077, LS 1267 (Star 86 in Table 7) and MF13 (we give the numbers in Moffat & Fitzgerald 1977 preceded by MF), which is outside our field. These data will be used in a forthcoming paper (Paper II) to determine the atmospheric parameters and abundances of these stars.

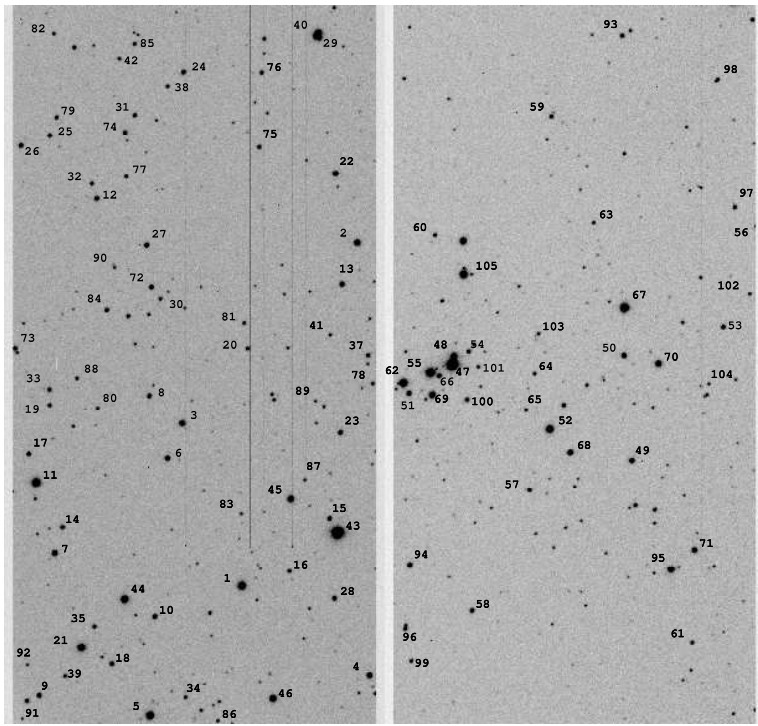

**Fig. 2.** Finding chart for stars in the field of Alicante 5. *XY* positions are listed in Table 6 where they are correlated to RA and DEC. The origin of coordinates is located at the bottom left corner of the image. The blank strip corresponds to the gap between the two CCDs of SUSI2. North is up and east is left.

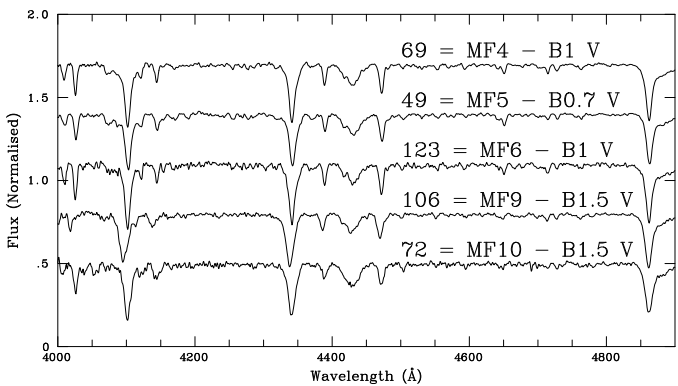

**Fig. 3.** Classification spectra of five early-type members of Pismis 11 within the field covered by our photometry. The numbering system from Moffat & Fitzgerald (1977) is also shown.

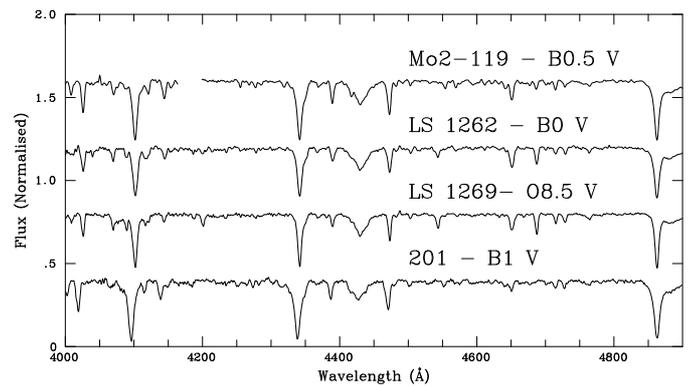

**Fig. 4.** Classification spectra of four early-type stars in the neighbourhood of Pismis 11. 201 lies to the South of the cluster and is likely to be an outlying member. The other three stars are located ∼ 15′ to the north of the cluster, not far from the small cluster Alicante 5.

Fig. 3 shows low-resolution spectra of 5 other members that fall within the field covered by our photometry. Their spectral types have been determined resorting to traditional criteria (Walborn & Fitzpatrick 1990) by comparison to standard stars in their digital library. In addition, we observed two relatively bright stars to the South of the cluster, whose 2MASS $(J - K_S)$ colours were similar to those of members. One of them, which we call star 201 (USNO B1.0 0399-0145111, $RA = 09^h15^m56.04^s$, $DEC = -50°05'20.4''$), turns out to be a likely member of spectral type B1 V, but we lack optical photometry for this object. We also took spectra of three catalogued early-type stars at some distance from the cluster. These objects are also listed in Table 7, while the spectra are displayed in Fig. 4.

### 3.1.2. Observational HR diagram

We start the photometric analysis by plotting the $V/(B - V)$ diagram for all stars in the field (see Fig. 5). We can observe that there is not a clear sequence, because the cluster is small and dispersed. There is strong contamination by field stars, forcing us to follow a careful analysis procedure.

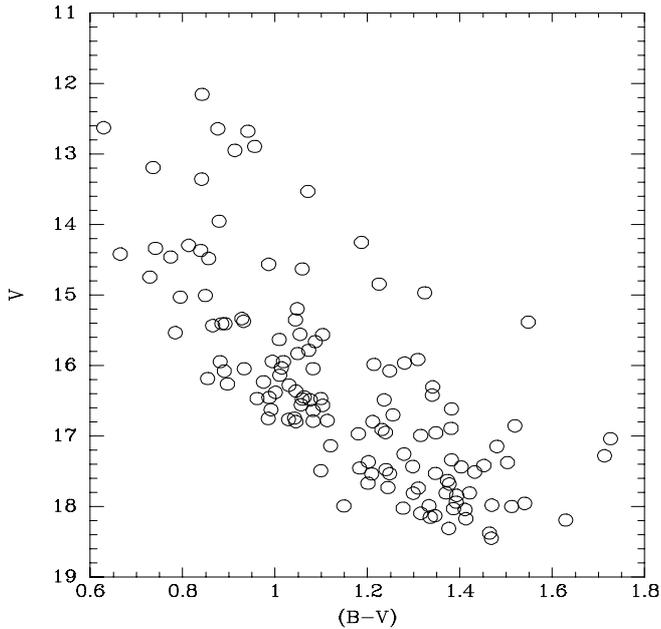

**Fig. 5.** $V/(B-V)$ diagram for all stars in the field of Pismis 11.

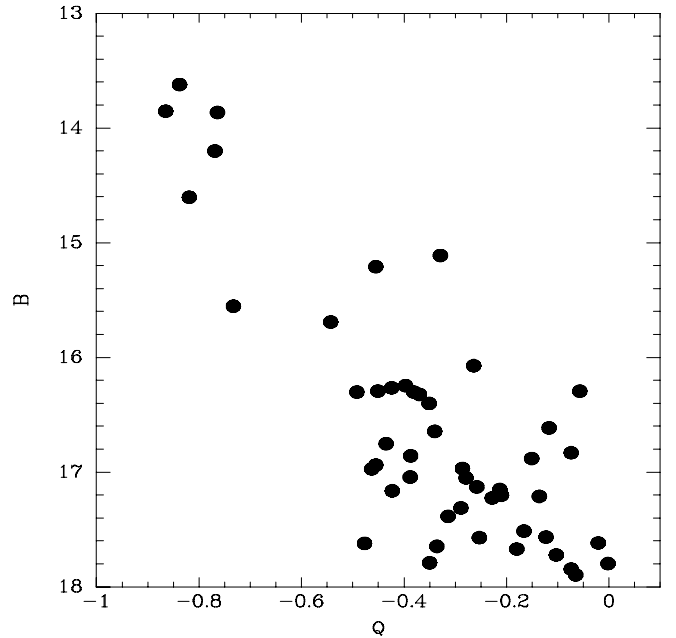

**Fig. 6.** $B - Q$ diagram for stars in the field of Pismis 11 for which we find $Q < 0$, corresponding to normal main-sequence B-type stars.

### 3.1.3. The reddening law

The first step is to determine whether the extinction law in the direction of the field is standard. We use the CHORIZOS ($\chi^2$ code for parametrized modeling and characterization of photometry and spectroscopy) code developed by Maíz-Apellániz (2004). This code fits synthetic photometry derived from the spectral distribution of a stellar model convolved with an extinction law to reproduce the observed magnitudes.

First, we select the stars with spectral types. In this sample, we have five stars falling within the region covered by our photometry (numbers 69, 49, 72, 123 and 106), for which we have $UBVRI$ photometry and $JHK_S$ photometry from the 2MASS catalogue. Star 86 (LS 1267) is saturated in our photometry. MF13 falls just outside the area covered by our photometry. For these two objects, we take $UBV$ photometry from Moffat & Fitzgerald (1977) and $JHK_S$ photometry from 2MASS. We have spectra of three more stars at some distance from the cluster, M02-119, LS 1262 and LS 1269. For these objects, we take $UBV$ photometry from Muzzio (1979) and $JHK_S$ photometry from 2MASS.

We use as input for CHORIZOS all the photometry available and the $T_{\rm eff}$ corresponding to the spectral type derived according to the calibrations of Morton & Adams (1968). The output of CHORIZOS is the value of $R$ and the excess $E(B-V)$. For all our stars, the preferred value of $R$ is close to 3.1, with little dispersion. This is taken as confirmation that the extinction law is standard in this direction. The average value of $E(B-V)$ for the five members in our photometric field corresponds to $1.17 \pm 0.06$.

### 3.1.4. The reddening-free $Q$ parameter and spectral types

The reddening-free $Q$ parameter allows a preliminary selection of early-type stars. The $Q$ parameter is defined as

$$Q = (U-B) - \frac{E(U-B)}{E(B-V)}(B-V). \qquad (1)$$

For a standard reddening law, $E(U-B)/E(B-V) = 0.72$ (Johnson & Morgan 1952). For the five stars in our dataset with spectra and $UBV$ photometry, we calculate the value of $E(U-B)/E(B-V)$, taking intrinsic colours from Morton & Adams (1968). We obtain an average value of 0.71, indicating that we can adopt the standard value for the $Q$ parameter and use it to select B-type stars (possible cluster members) using the calibration of $Q$ against spectral type (Johnson & Morgan 1952). As all B-type stars are expected to have a negative $Q$ parameter, we select only stars with $Q < 0$. Considering the young age of the cluster implied by the presence of O-type stars, stars later than mid A spectral type must be still contracting towards the ZAMS. Moreover, from the observed magnitudes of the brightest cluster members and their spectral types, we would expect any possible main-sequence members later than ~A0 to be fainter than our faint magnitude limit. Therefore we are not leaving out any likely member by selecting only B-type stars. The stars that could have B spectral type based on their $Q$ parameter are shown in Fig. 6.

In order to confirm the membership of these possible B-type stars, we need a more accurate spectral typing, based only on photometry. Again, we make use of the fact that the reddening is standard in this direction. In Fig. 7, we plot the $(U-B)/(B-V)$ diagram for stars with $Q < 0$. We draw the unreddened main sequence and the main sequence reddened by the average value $E(B-V) = 1.2$ (the value from Moffat & Fitzgerald 1977 is identical, within the errors, to that derived here). We also indicate reddening lines with slope 0.72 for each spectral type, according to the calibration of Johnson & Morgan (1952). We can check that stars with known spectral types ($\approx$B1 in all cases) are located close to the correct line. Now, we can assign spectral types for each star according to its location in this diagram. The approximate spectral types assigned are listed in Table 8. The $Q$ parameter has been calibrated mostly using stars less reddened than those in Pismis 11. The calibration does not take into account bandpass effects, which may be important at high reddening. For this reason, the pseudo spectral types derived should only be considered approximations, which will be used as input for CHORIZOS. The code takes into account bandpass effects.

**Table 7.** Spectral types and photometry for stars in the region. The top panel shows likely members of Pismis 11. The second panel contains likely members of Alicante 5. The bottom panel displays stars outside the clusters.

| Star | Spectral Type | RA(J2000) | DEC(J2000) | $V$ | $(B-V)$ | $(U-B)$ |
|---|---|---|---|---|---|---|
| 69 | B1 V | 09 15 54.046 | −50 00 26.67 | 12.681 | 0.941 | −0.161 |
| 49 | B0.7 V | *09 15 54.570 | −50 00 03.90 | 12.898 | 0.956 | −0.177 |
| LS 1267 (86) | O8 V | 09 15 52.787 | −50 00 43.83 | $11.07^a$ | $1.0^a$ | $-0.19^a$ |
| 106 | B1.5 Vn | 09 15 51.763 | −50 01 41.63 | 13.360 | 0.841 | −0.164 |
| 123 | B1 V | 09 15 52.053 | −50 02 38.93 | 12.952 | 0.913 | −0.107 |
| HD 80077 | B2 Ia$^+$ | 09 15 54.79 | −49 58 24.6 | $7.58^a$ | $1.37^a$ | $0.20^a$ |
| MF13 | B0.2 V | 09 15 31.08 | −50 02 21.5 | $12.03^a$ | $0.85^a$ | $-0.24^a$ |
| A47 | B0.7 V | 09 16 29.71 | −49 42 49.6 | 13.259 | 1.204 | 0.164 |
| A55 | B3 III? | 09 16 28.85 | −49 42 53.0 | 14.419 | 1.247 | 0.403 |
| A62 | B1.5 V | 09 16 27.47 | −49 42 59.0 | 15.130 | 1.222 | 0.458 |
| MO 2-119 | B0.5 V | 09 15 33.7 | −49 48 06.0 | $11.74^c$ | $1.12^c$ | - |
| LS 1262 | B0 V | 09 15 05.17 | −49 44 13.9 | $11.17^b$ | $0.81^b$ | $-0.30^b$ |
| LS 1269 | O8.5 V | 09 15 56.60 | −49 45 02.9 | $10.85^b$ | $0.92^b$ | $-0.19^b$ |

$^{(a)}$ Photometry for MF13, LS 1267 and HD 80077 from Moffat & Fitzgerald (1977).
$^{(b)}$ Photometry for LS 1262, LS 1269 from Muzzio (1979).
$^{(c)}$ Photometry for M02-119 from Muzzio & Orsatti (1977).

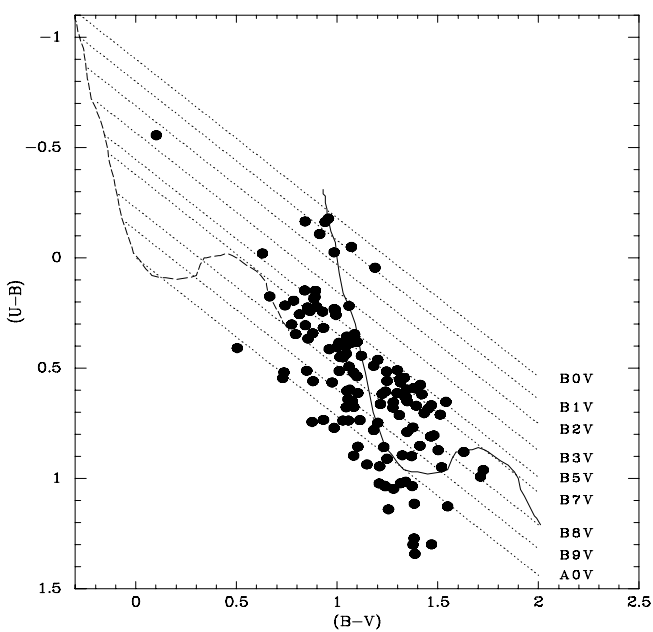

**Fig. 7.** $(B-V)/(U-B)$ diagram for stars in Pismis 11. The dashed curve shows the position of the unreddened main sequence. The solid line is the main sequence reddened by the average cluster excess $E(B-V) = 1.2$. The straight lines are the reddened locations of stars of a given spectral type.

We select as candidate members those objects forming a sequence in the $V-(B-V)$ diagram, rejecting those stars whose magnitude and approximate spectral type seemed completely incompatible with membership in the sequence.

### 3.1.5. Determination of the distance

After this last selection of candidate members, we use the CHORIZOS code again. In this case, we use as input data the $UBVRIJHK_s$ magnitudes (our photometry + 2MASS), the $T_{\rm eff}$ corresponding to the spectral types derived from the photometry in the previous section (according to the calibration of Morton & Adams 1968) and $\log g$ appropriate for a main-sequence star. The output from CHORIZOS are individual values of the excess $E(B-V)$ and the dereddened $V$ magnitudes. In a first analysis, we reject stars with very low $E(B-V)$ ($< 1$) as foreground stars. We obtain the diagram shown in Fig. 8. In this diagram we also plot two different observational ZAMS, those from Mermilliod (1981) and from Schmidt-Kaler (1982). The top of the main sequence gives a good fit to the Schmidt-Kaler ZAMS if a distance modulus $DM = 12.8$ is adopted. The lower part of the sequence deviates from the ZAMS. A better fit is obtained if we use the Mermilliod ZAMS. Some stars fall below both ZAMS. Their $E(B-V)$ is in all cases much higher than the expected average for the cluster $E(B-V) = 1.2$. Pismis 11 is located at galactic longitude $l = 271°$, in a region where disk tracers are common at different distances (Brand & Blitz 1993, Russeil 2003) and therefore a background or foreground population is not unexpected. In view of this, we conclude that these faint objects are background stars and decide to keep as likely members only objects with $E(B-V)$ in the interval 1.0–1.6 mag. We cannot completely discard the possibility that these objects with higher reddening are also members and that the dereddening procedure used fails at the highest reddenings. But this would not have any impact on the distance determination.

It is also possible that some of the stars at the top of the main sequence may be binaries or multiple systems. Because of all these uncertainties, we adopt a conservatively high ±0.3 error for our $DM$ and accept $DM = 12.8 \pm 0.3$, based on 43 very likely members. This $DM$ corresponds to a distance of $3.6^{+0.6}_{-0.4}$ kpc (see Fig. 8).

We plot individual values of $E(B-V)$ for members in Fig. 9. There is some dispersion and values are not strongly correlated with spatial location. The claim by previous authors that the reddening is higher in the subfield next to HD 80077 cannot be ascertained, though a number of objects with high reddening lie in the immediate vicinity of the supergiant.

### 3.2. The new cluster Alicante 5

Fig. 8 shows that some early stars outside the cluster fit the cluster main sequence well. Therefore they must be located at the same distance as Pismis 11. While inspecting maps of this region for other catalogued OB stars in the area, we noticed the presence of a small group of stars resembling a stellar cluster, ∼ 18′ north and slightly to the west of Pismis 11.

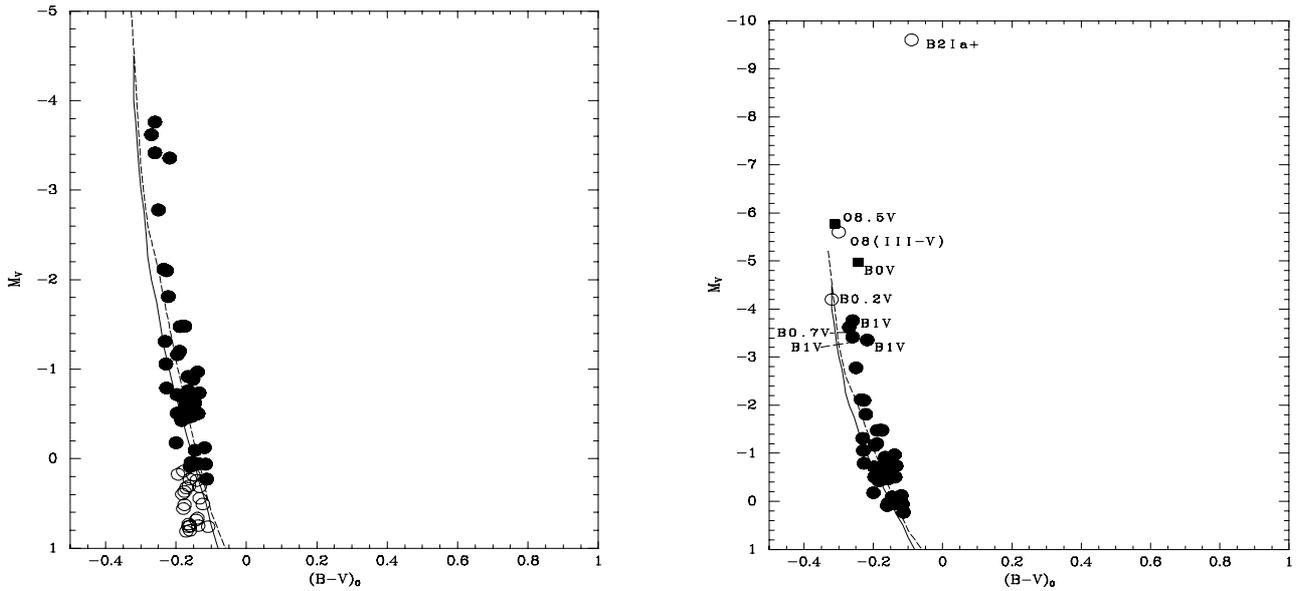

**Fig. 8.** Left: Dereddened $M_V/(B-V)$ diagram for early type stars in the field of Pismis 11. The solid line is the ZAMS (from Mermilliod 1981 and the dash line is the ZAMS (from Schmidt-Kaler 1982). Open circles represent stars that do not fit any of the ZAMS. As all of them have higher reddening than cluster members, we take them for background stars. Right: Dereddened $M_V/(B-V)$ diagram for likely members of Pismis 11 and other catalogued OB stars in the neighbourhood. Open circles and filled squares correspond to stars with photometry from Moffat & Fitzgerald (1977) and Muzzio (1979) or Muzzio & Orsatti (1977), respectively.

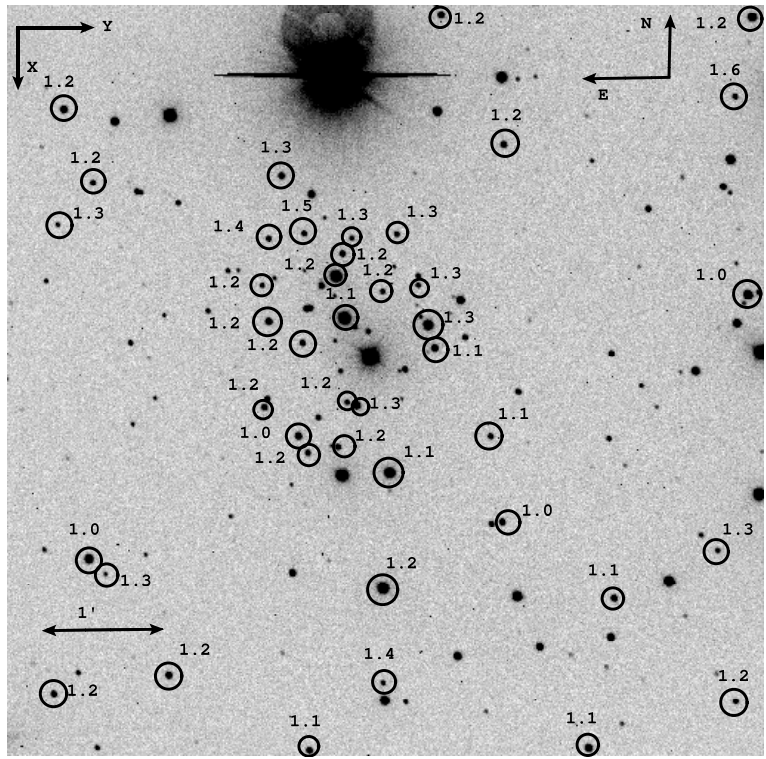

**Fig. 9.** Map showing all the likely members of Pismis 11 selected from our photometry with their derived individual reddenings. Stars towards the north of the cluster seem to have on the average higher reddenings than those to the south, though the difference is not statistically significant.

### 3.2.1. Spectroscopy

We took spectra of the 3 brightest stars in the group, finding that they are all B-type stars. The spectra are shown in Fig. 10.

The spectrum of star A47 is of good quality and we derive a spectral type B0.7 V. The spectra of the two other stars (A55

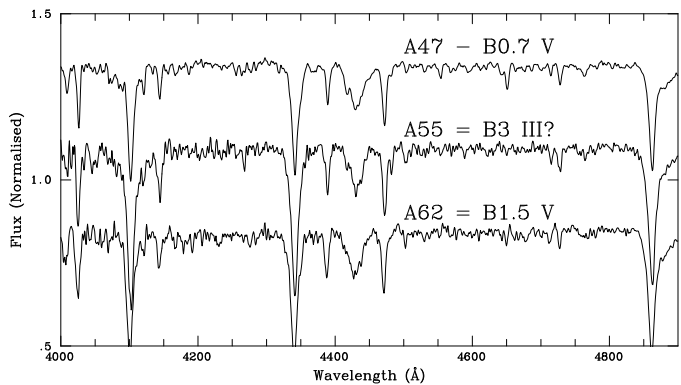

**Fig. 10.** Classification spectra of the three brightest stars in the new cluster Alicante 5. The spectra of A55 and A61 have low SNR and the spectral types derived are tentative.

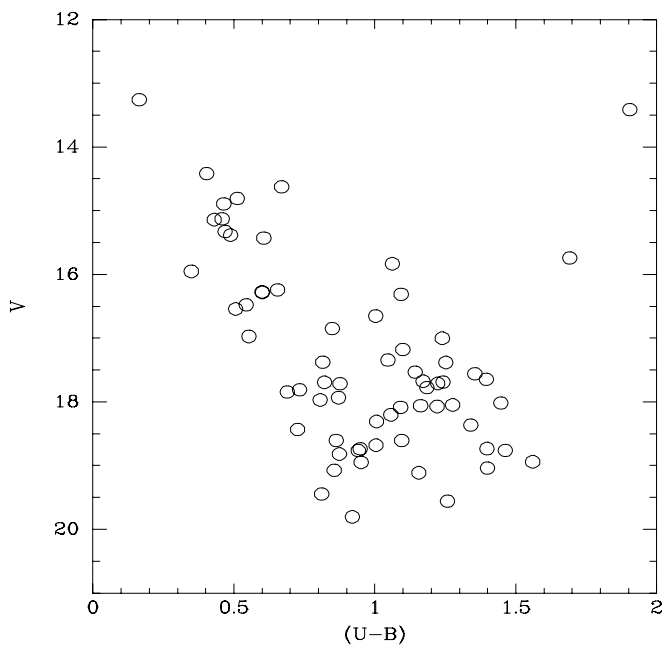

**Fig. 11.** Observational $V/(U-B)$ diagram for all stars in the field of Alicante 5.

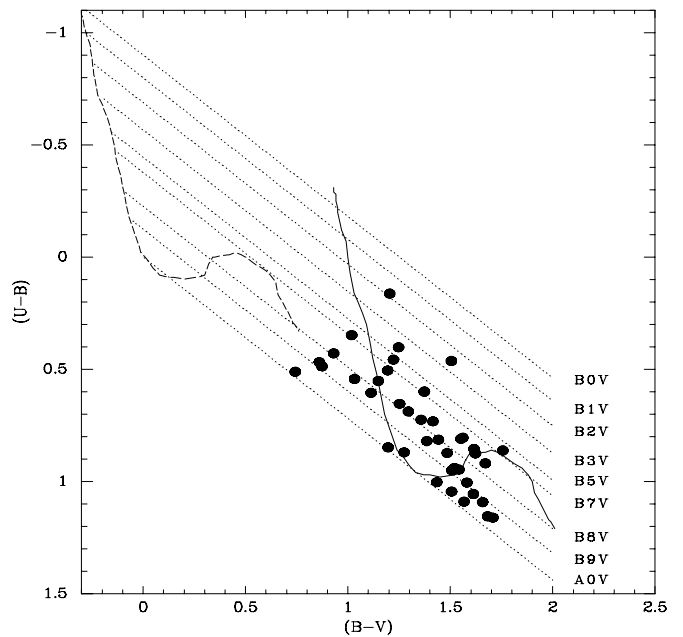

**Fig. 12.** $(B-V)/(U-B)$ diagram for stars in Alicante 5. The dashed curve shows the position of the unreddened main sequence. The solid line is the main sequence reddened by the average cluster excess $E(B-V) = 1.2$. The straight lines are the reddened locations of stars of a given spectral type.

and A62) have low signal to noise ratio and are more difficult to classify. Based on the possible presence of a strong C II 4267Å line, A55 is tentatively classified as B3 III, but this is uncertain. Star A62 is B1.5 or B2 and seems to be on the main sequence.

### 3.2.2. Photometric analysis

In order to investigate the existence of a new cluster and to determine its distance, we obtained $UBVRI$ photometry for 105 stars around the stellar concentration. Photometric values for these objects are found in Table 5.

With these data, we carry out an analysis similar to that done for Pismis 11. First, we plot the $(U-B)/V$ diagram (see Fig. 11). From the start, we can see a likely sequence but it is not very clear, and we need to investigate how many stars are likely to belong to this concentration. For that aim, we build the $(B-V)/(U-B)$ diagram (see Fig. 12) and assign an approximate spectral type to each star (see Table 9), again assuming a standard reddening law. As in the previous case, those stars whose spectral types and positions in the $(U-B)/V$ diagram are incompatible are removed from the sample. With the stars left, we create an input file for CHORIZOS, which is used to determine their individual values of $E(B-V)$ and $A_V$ (the values are listed in Table 9). With the dereddened values of $V$ and $(B-V)$ (see Table 9), we plot the corresponding diagram. The photometric sequence can be fit to the ZAMS using a $DM = 12.8$, identical to the value used for Pismis 11 (see Fig. 13). In view of this, we confirm the fact that this grouping is an uncatalogued young open cluster (which we provisionally call Alicante 5). We also confirm that it is at the same distance as Pismis 11 and its neighbouring OB stars. The selection procedure leaves 22 likely members for Alicante 5, all of which should be B-type stars.

## 4. Discussion

### 4.1. Cluster parameters

Based on a ZAMS fit to the dereddened colour-magnitude diagram for 43 likely members, we find a true distance modulus to Pismis 11 $DM = 12.8 \pm 0.3$, corresponding to $d = 3.6^{+0.6}_{-0.4}$ kpc. The reddening in the area, though patchy, follows a standard $R = 3.1$ law. The average colour excess for likely members is $E(B-V) = 1.2$, with most values in the 1.1–1.3 range and with few outliers.

These results are in good agreement with those of Moffat & Fitzgerald (1977), who were able to obtain an accurate distance from a very small number of members because the reddening follows the standard law in this area of the sky.

The age of the cluster cannot be estimated from the data available, as the only evolved star is HD 80077. The presence of unevolved B-type stars sets the age to $\lesssim 10$ Myr, while the fact that LS 1267 is an O8 star still close to the ZAMS implies $\lesssim 5$ Myr. If HD 80077 is a cluster member, Carpay et al. (1989) estimate its age to be 3.0 Myr from its position in the theoretical HR diagram.

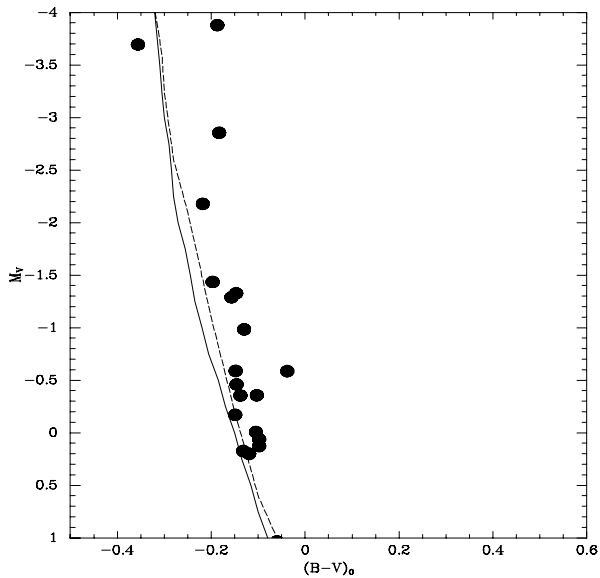

**Fig. 13.** Dereddened $M_V/(B-V)$ diagram for likely members of Alicante 5. The solid line shows the ZAMS from Mermilliod (1981) and the dash line the ZAMS from Schmidt-Kaler (1982). The position of the bluest star (A55) has been calculated assuming a main sequence star, while the spectrum (see Fig. 10) suggests it may be a giant star.

### 4.2. Association with other stars around the cluster

Several catalogued OB stars can be found in the neighbourhood of Pismis 11. As shown in Fig. 8, these objects fit the cluster main-sequence well, indicating a common distance. With the data available, the exact size of the cluster cannot be assessed. Cluster members are found outside the area covered by our photometry (201, MF13). However, stars like LS 1262 and LS 1269 are simply too far away from the cluster ($\sim 15'$) to be considered members. In this respect, our finding of a second smaller cluster in the vicinity of these stars, at $\sim 18'$ from Pismis 11, which fits the same main sequence and has a population of B-type stars scattered around it, suggests that we are seeing an extended OB association, containing some small clusters. At a distance of 3.6 kpc, the distance between the two clusters corresponds to $\approx 18$ pc, compatible with the typical size of an association.

We note that some of the stars in Fig. 8 (particularly LS 1262 and LS 1269) have absolute magnitudes somewhat higher than expected for their spectral types. The difference can be explained by a binary nature of the stars (a companion of the same spectral type adds 0.7 mag), but the possibility of a small population at a shorter distance than Pismis 11 cannot be entirely ruled out.

### 4.3. Membership of HD 80077

One of the main aims of this work is to determine if HD 80077 is a member of Pismis 11. Though HD 80077 is slightly outside the core of the cluster ($\sim 2'$), there are other massive members at larger distances. Carpay et al. (1991) suggested that this object was extremely luminous, as it appeared much more lumi-nous than the B2 Ia supergiant $\chi^2$ Ori, which is estimated to have $\log(L/L_\odot) = 5.65$ (Crowther et al. 2006).

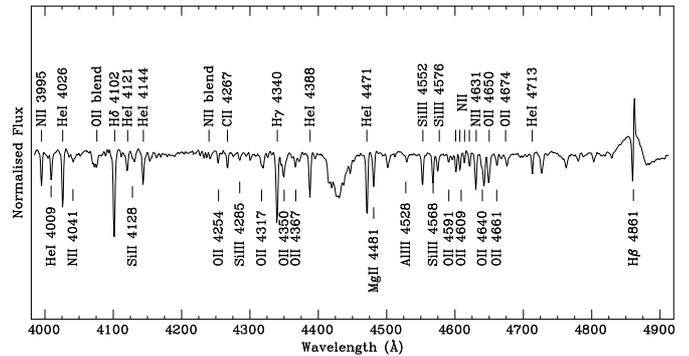

**Fig. 14.** Blue spectrum of HD 80077, taken in 2003, showing features indicative of a very high luminosity, such as the broad emission peak in H$\beta$, the prominence of Mg II 4881Å for the B2 Ia and the extreme weakness of He I lines when compared to metallic features.

**Table 10.** Proper motions for stars in the region. The top panel shows likely members of Pismis 11. The bottom panel displays stars outside the cluster. All measurements are in mas yr$^{-1}$.

| Star | Tycho2 | | UCAC2 | |
|---|---|---|---|---|
| | P.M. ($\alpha \cos \delta$) | P.M. ($\delta$) | P.M. ($\alpha$) | P.M. ($\delta$) |
| 69 = MF4 | – | – | –4.6 ± 2.5 | 10.6 ± 2.3 |
| 86 = LS 1267 | –1.7 ± 2.1 | 4.8.0 ± 2.0 | –1.2 ± 1.4 | 7.0 ± 2.6 |
| 106 = MF9 | – | – | –19.9 ± 4.7 | 8.6 ± 4.7 |
| 123 = MF6 | – | – | –9.2 ± 2.5 | 3.9 ± 2.3 |
| HD 80077 | –3.8 ± 1.1 | 4.0 ± 1.0 | –7.3 ± 0.7 | 4.8 ± 0.6 |
| S13 | | | –10.0 ± 1.4 | 11.8 ± 1.9 |
| M02-119 | –3.4 ± 2.4 | 11.5 ± 2.3 | –3.0 ± 3.5 | 8.6 ± 1.4 |
| LS 1262 | –7.8 ± 2.2 | 4.0 ± 2.0 | –8.5 ± 1.4 | 4.8 ± 1.4 |
| LS 1269 | –2.2 ± 2.5 | 0.5 ± 2.4 | –1.9 ± 1.5 | 2.5 ± 1.9 |

#### 4.3.1. Kinematics

Knoechel & Moffat (1982) measured the radial velocity of HD 80077 and found important variations. Though they could not discard the possibility that they were simply due to wind variability, they slightly favoured the idea that it was a spectroscopic binary, perhaps with a compact companion. The systemic radial velocity is very low, and compatible with its position in the Galaxy. In this direction, objects at different heliocentric distances have similar galactocentric distance and similarly low $v_{\rm LSR} \lesssim 10\,{\rm km\,s^{-1}}$ (e.g., Brand & Blitz 1993, Russeil 2003).

HD 80077 is bright enough to have an accurate measurement of its proper motion in the Hipparchos catalogue, namely, P.M.($\alpha \cos \delta$) = $-4.7 \pm 0.7$ mas yr$^{-1}$, P.M.($\delta$) = $4.8 \pm 0.6$ mas yr$^{-1}$. Unfortunately, no other stars in the area are bright enough to have *Hipparchos* measurements. However, several of them are in the *Tycho2* catalogue and all the brightest members of Pismis 11 have measurable proper motion in the UCAC2 catalogue. The values of these proper motions are reported in Table 10. As seen, all stars in the area except LS 1269 have proper motions which appear compatible within the (admittedly very high) errors and hence it is not unreasonable to assume that they are close to the much more accurate values for HD 80077. This is not surprising, as these proper motions are typical of OB stars in this direction at a range of heliocentric distances.

The tangential velocity component due to the galactic orbital motion is very small for stars close to the Sun and increases with distance, as the angle with the Sun – Galactic Centre vector increases. At $l = 271°$ and a distance of 3.6 kpc, this angle is $\sim 23°$. For an assumed circular motion with $v_{orb} = 220$ km s$^{-1}$, the tangential component would be $v_{tan} \approx 90$ km s$^{-1}$. The total tangential velocity of HD 80077, derived from its proper motions at a distance of 3.6 kpc would be $v_{tan} = 91 \pm 23$ km s$^{-1}$. Unfortunately, this excellent agreement does not give us any information about the distance to HD 80077, because at $l \approx 270°$, a similar agreement can be obtained at *any distance*. It tells us, however, that HD 80077 has no measurable peculiar velocity and hence it is not a runaway. The same can be said for all other stars in the area, except LS 1269.

### 4.3.2. Reddening and luminosity

Knoechel & Moffat (1982) analysed the interstellar Ca II lines in the spectrum of HD 80077, coming to the conclusion that the extinction of the star was mostly caused by dust within 2 kpc of the Sun. The colour excess of HD 80077, $E(B - V) = 1.46$, is slightly higher than the average for members, but inside the range considered typical (1.0,1.6). The reddening in this area is patchy, and some members lying close to HD 80077 also have higher than average values of $E(B - V)$, even though there does not seem to be any clear gradient across the face of the cluster (see Fig. 9).

The position of HD 80077 in the HR diagram of the cluster fits well with its photometric values and spectral type. Besides, the presence of moderately massive stars still on the ZAMS allows the progenitor of HD 80077 to have originally had $M_* \gtrsim 40 M_\odot$. The gap of 4 mag between HD 80077 and the next brightest star, LS 1267, is larger than usually seen in open clusters with supergiants. However, given the low number of massive stars, its significance is difficult to assess.

The strongest argument used against the membership of HD 80077 (e.g., Knoechel & Moffat 1982) comes from the lack of strong evidence for heavy mass loss. If HD 80077 is a member of Pismis 11, it must have $M_V = -9.5$ corresponding to a $M_{bol} \approx -10.5$, assuming a bolometric correction similar to other B2 Ia supergiants (Humphreys & McElroy 1984). This would mean that it is one of the most luminous stars in the Galaxy, as discussed by Carpay et al. (1989; 1991), lying above the empirical Humphreys-Davidson limit (Humphreys & Davidson 1994), where stars are supposed to be unstable.

Moreover, Knoechel & Moffat (1982) argued that the lightcurve of HD 80077 only showed small variability, typical of luminous supergiants. On the other hand, van Genderen et al. (1992), observing the star for a much longer timespan, found larger light variations and concluded that HD 80077 was perhaps a luminous blue variable in quiescence. The few spectra of the star published suggest that it has not displayed any sign of spectral variability over 30 years.

There is, however, one obvious peculiarity in the spectrum of HD 80077. Knoechel & Moffat (1982) observed a sharp P-Cygni profile in the H$\beta$ line on top of a weak broad emission feature on photographic Coudé plates taken in February – March 1977. This broad feature is also seen in our spectrum (Fig. 14). Though the intensity of the broad feature is not easy to assess in this spectrum, because of the nearness of the 4885Å diffuse interstellar band and the edge of the CCD (affecting the normalisation), it seems to be present in all the spectra of HD 80077. This unusual shape of H$\beta$ is highly reminiscent of that seen in the quiescent LBV HR Car and the LBV candidate HD 168607 (e.g., Walborn & Fitzpatrick 2000). On the other hand, other early B hypergiants, like $\zeta^1$ Sco or Wray 977, present strong P-Cygni profiles in He I lines (e.g., Kaper et al. 2006), which are not seen in HD 80077. The main difference between the spectrum of HR Car and that of HD 80077 (in addition to the later spectral type of HR Car) is the presence of many Fe II and [Fe II] emission lines in the former (Machado et al. 2002). In addition, the H$\alpha$ P-Cygni of HD 80077 seen in our 2003 spectra peaks 2.6 times above the continuum level, while it only reaches $\sim 1.2$ times the continuum in a sample of B1 – B2 Ia supergiants. There is thus evidence for a higher mass loss rate than in typical B2 Ia supergiants and spectral similarities to a known LBV.

Based on the analysis of the interstellar lines, Knoechel & Moffat (1982) suggested that HD 80077 had to be at a distance $d \gtrsim 2$ kpc. Taking into account that the star is not a runaway, there is no stellar association to which it could belong if it is not a cluster member. The association Vela OB1 lies at $d = 1.9$ kpc (Humphreys 1978), but it is $> 5°$ apart in the sky. Even at this shorter distance, HD 80077 would have $M_V = -8.6$, and still be a B2 Ia$^+$ hypergiant according to the calibration of Humphreys & McElroy (1984). As a comparison, $\zeta^1$ Sco has $M_V = -8.8$ (Crowther et al. 2006). The distance to HR Car is not well established. Accepting the high value of van Genderen et al. (1991), it has $M_{bol} = -8.9$. For a B3 Ia spectral type (Walborn & Fitzpatrick 2000) this imply a $M_V = -7.8$.

If HD 80077 is a member of Pismis 11, it will then be brighter than stars with similar spectra. In order to resolve the issue of membership, in Paper II we will carry out a detailed analysis of echelle spectra of HD 80077 and the brightest OB members of Pismis 11, which will be used to derive their stellar parameters and surface abundances. With this, we expect to assess if HD 80077 can really be as luminous as needed for it to be a member of Pismis 11 or rather is more compatible with a foreground object.

## 5. Conclusions

From a careful analysis of individual reddenings, we find a standard reddening law to the open cluster Pismis 11. Pending a decision on the membership of HD 80077, the cluster contains one O-type star and several early B-type stars. We derive a distance of 3.6 kpc, in good agreement with the work of Moffat & Fitzgerald (1977). Around the cluster, we find a number of early-type stars at the same distance, and a small open cluster, which we call Alicante 5, located $\sim 18'$ away from Pismis 11, which also shares the same distance modulus. We find no strong reason to doubt the membership of HD 80077 in Pismis 11, but then it would have to be one of the most luminous stars in the Galaxy and perhaps more extreme spectral features should be expected. In a second paper, we will carry out a detailed analysis of intermediate and high resolution spectra of HD 80077 and the three brightest OB members of Pismis 11, and a comparison to the blue hypergiant $\zeta^1$ Sco, in order to assess the luminosity of HD 80077.

*Acknowledgements.* We thank the referee, Nolan Walborn, for many helpful suggestions. This research is partially supported by the MEC under grants AYA2005-00095 and CSD2006-70. This research has made use of the Simbad database, operated at CDS, Strasbourg (France). This publication makes use of data products from the Two Micron All Sky Survey, which is a joint project of the University of Massachusetts and the Infrared Processing and Analysis Center/California Institute of Technology, funded by the National Aeronautics and Space Administration and the National Science Foundation. These observations have been funded by the Optical Infrared Coordination Network (OPTICON), a major international collaboration supported by the Research

## References

Brand, J., & Blitz, L. 1993, A&A, 275, 67
Carpay, J., de Jager, C., Nieuwenhuijzen, H., & Moffat, A. 1989, A&A, 216, 143
Carpay, J., de Jager, C., & Nieuwenhuijzen, H. 1991, A&A, 248, 475
Crowther, P.A., Lennon, D.J., & Walborn, N.R. 2006, A&A, 446, 279
Draper, P.W., Taylor, M., & Allan, A. 2000, Starlink User Note 139.12, R.A.L.
Howarth, I., Murray, J., Mills, D., & Berry, D.S. 1998, Starlink User Note 50.21, R.A.L.
Humphreys, R.M. 1978, ApJS 38, 309
Humphreys, R.M.,& McElroy, D.B. 1984, ApJ, 284, 565
Humphreys, R.M., & Davidson, K. 1994, PASP 106, 1025
Johnson, H.L., & Morgan, W.W. 1952, ApJ, 117, 313
Kaper, L., van den Meer, A., & Najarro, F. 2006, A&A, 457, 595
Knoechel, G., & Moffat, A. F. J. 1982, A&A, 110, 263
Kudritzki, R.P., & Puls, J. 2000, ARA&A, 38, 613
Kudritzki, R. P., Bresolin, F., & Przybilla, N. 2003, ApJ, 582, L83
Landolt, A.U. 1992, AJ, 104, 340
Machado, M.A.D., de Araújo, F.X., Pereira, C.B., & Fernandes, M.B. 2002, A&A 387, 151
Maíz-Apellániz, J. 2004, PASP, 116, 859
Martins, F., Schaerer, D., & Hillier, D.J. 2005, A&A, 436, 1049
Mermilliod, J.C. 1981, A&A, 97, 235
Moffat, A. F. J., & Fitzgerald, M. P. 1977, A&A, 54, 263
Morton, D.C., & Adams, T.F. 1968, ApJ, 151, 611
Muzzio, J. C., & Orsatti, A. M. 1977, AJ, 82, 474
Muzzio, J. C. 1979, AJ, 84, 639
Russeil, D. 2003, A&A, 397, 133
Shortridge, K., Meyerdicks, H., Currie, M., et al. 1997, Starlink User Note 86.15, R.A.L.
Skrutskie, M.F., Cutri, R.M., & Stiening, R. 2006, AJ, 131, 1163
Schmidt-Kaler, T. 1982, Landolt-Börnstein, N.S. VI, 2b
Stetson, P. B. 1987, PASP, 99, 191
van Genderen, A.M. 2001, A&A, 366, 508
van Genderen, A.M., Robijn, F.H.A., van Esch, B.P.M., & Lamers, H.J.G.L.M. 1991, A&A, 246, 407
van Genderen, A.M., van den Bosch, F.C., Dessing, F., et al. 1992, A&A, 264, 88
Walborn, N.R., & Fitzpatrick, E.L. 1990, PASP, 102, 379
Walborn, N.R., & Fitzpatrick, E.L. 2000, PASP, 112, 50

12             A. Marco and I. Negueruela: Pismis 11 and HD 80077**Table 2.** (X, Y) position on the map of stars with photometry in the field. USNO-B1 identification for these stars and their coordinates.

| Number | X (Pixels) | Y (Pixels) | RA (J2000) | DEC (J2000) | Name (USNO-B1.0) |
|---|---|---|---|---|---|
| 1 | - | - | 09:16:00.46 | -49:58:16.2 | 2MASS |
| 2 | 13.28 | 587.48 | 09:15:49.431 | -49:57:53.94 | 0400-0143996 |
| 3 | - | - | 09:15:44.67 | -49:57:47.0 | 2MASS |
| 4 | - | - | 09:15:45.50 | -49:57:47.4 | 2MASS |
| 5 | 16.29 | 1008.17 | 09:15:33.591 | -49:57:54.62 | 0400-0143756 |
| 6 | 1.50 | 580.50 | 09:16:11.94 | -49:58:19.9 | 2MASS |
| 7 | - | - | - | - | |
| 8 | 58.59 | 703.92 | 09:15:44.978 | -49:58:11.22 | 0400-0143930 |
| 9 | - | - | 09:16:11.94 | -49:58:19.9 | 2MASS |
| 10 | 96.88 | 714.98 | 09:15:44.595 | -49:58:24.14 | 0400-0143924 |
| 11 | 97.80 | 669.23 | 09:15:46.233 | -49:58:25.79 | 0400-0143947 |
| 12 | 100.22 | 691.10 | 09:15:45.396 | -49:58:24.35 | 0400-0143939 |
| 13 | - | - | - | - | |
| 14 | 123.71 | 985.26 | 09:15:34.237 | -49:58:34.63 | 0400-0143770 |
| 15 | 141.01 | 77.06 | 09:16:08.498 | -49:58:42.87 | 0400-0144266 |
| 16 | 143.49 | 582.09 | 09:15:49.558 | -49:58:45.77 | 0400-0143998 |
| 17 | 149.50 | 219.50 | 09:16:03.078 | -49:58:45.45 | 0400-0144200 |
| 18 | 157.11 | 146.02 | 09:16:05.890 | -49:58:49.25 | 0400-0144239 |
| 19 | 189.54 | 672.09 | 09:15:46.125 | -49:58:58.44 | 0400-0143945 |
| 20 | 209.62 | 978.78 | 09:15:34.639 | -49:59:04.79 | 0400-0143778 |
| 21 | 211.53 | 791.55 | 09:15:41.647 | -49:59:06.03 | 0400-0143882 |
| 22 | 224.50 | 715.50 | 09:15:44.445 | -49:59:10.88 | 0400-0143919 |
| 23 | 231.59 | 371.62 | 09:15:57.397 | -49:59:14.90 | 0400-0144122 |
| 24 | 233.33 | 913.69 | 09:15:37.038 | -49:59:13.46 | 0400-0143817 |
| 25 | 227.50 | 550.50 | 09:15:50.660 | -49:59:12.75 | 0400-0144017 |
| 26 | 241.21 | 116.61 | 09:16:06.991 | -49:59:19.31 | 0400-0144252 |
| 27 | 246.87 | 968.78 | 09:15:34.915 | -49:59:18.46 | 0400-0143783 |
| 28 | 253.50 | 175.50 | - | - | - |
| 29 | 254.00 | 176.59 | - | - | - |
| 30 | 255.51 | 978.34 | 09:15:34.352 | -49:59:24.57 | 0400-0143775 |
| 31 | 257.75 | 411.97 | 09:15:55.839 | -49:59:23.57 | 0400-0144109 |
| 32 | 269.16 | 231.68 | 09:16:02.643 | -49:59:28.78 | 0400-0144191 |
| 33 | 277.19 | 867.39 | 09:15:38.792 | -49:59:27.97 | 0400-0143838 |
| 35 | 289.32 | 429.70 | 09:15:55.082 | -49:59:35.30 | 0400-0144086 |
| 36 | 298.67 | 68.87 | 09:16:08.701 | -49:59:40.23 | 0400-0144269 |
| 38 | 310.56 | 402.15 | 09:15:56.251 | -49:59:42.84 | 0400-0144113 |
| 39 | 311.31 | 527.62 | 09:15:51.522 | -49:59:43.03 | 0400-0144026 |
| 40 | 316.61 | 466.48 | 09:15:53.815 | -49:59:45.22 | 0400-0144050 |
| 41 | 317.24 | 354.13 | 09:15:58.018 | -49:59:45.87 | 0400-0144126 |
| 42 | 339.05 | 453.99 | 09:15:54.289 | -49:59:53.41 | 0400-0144061 |
| 43 | 352.18 | 986.81 | 09:15:34.283 | -49:59:56.24 | 0400-0143772 |
| 44 | 361.31 | 312.93 | 09:15:59.550 | -50:00:01.50 | 2MASS |
| 45 | 356.50 | 523.89 | 09:15:51.679 | -49:59:59.46 | 0400-0144030 |
| 46 | 356.64 | 854.88 | 09:15:39.206 | -49:59:57.74 | 0400-0143852 |
| 47 | 361.87 | 299.04 | 09:16:00.070 | -50:00:01.80 | 2MASS |
| 48 | 361.82 | 402.45 | 09:15:56.167 | -50:00:02.22 | 0399-0145113 |
| 49 | 369.60 | 445.31 | 09:15:54.570 | -50:00:03.90 | 2MASS |
| 50 | 369.30 | 556.44 | 09:15:50.430 | -50:00:03.30 | 2MASS |
| 51 | 372.00 | 361.98 | 09:15:57.726 | -50:00:05.70 | 0399-0145138 |
| 52 | 375.77 | 53.04 | 09:16:09.276 | -50:00:08.05 | 0399-0145260 |
| 53 | 381.31 | 556.32 | 09:15:50.410 | -50:00:07.90 | 2MASS |
| 54 | 381.60 | 345.50 | 09:15:58.260 | -50:00:08.70 | 0399-0145151 |
| 55 | 380.26 | 184.48 | 09:16:04.336 | -50:00:09.06 | 0399-0145213 |
| 56 | 381.86 | 425.66 | 09:15:55.310 | -50:00:08.20 | 2MASS |
| 57 | 389.76 | 508.28 | 09:15:52.218 | -50:00:11.45 | 0399-0145044 |

**Table 2.** (X, Y) position on the map of stars with photometry in the field. USNO-B1 identification for these stars and their coordinates.

| Number | X (Pixels) | Y (Pixels) | RA (J2000) | DEC (J2000) | Name (USNO-B1.0) |
|---|---|---|---|---|---|
| 58 | 395.59 | 443.12 | 09:15:54.552 | -50:00:12.52 | 0399-0145092 |
| 59 | 402.14 | 613.61 | 09:15:48.314 | -50:00:15.54 | 0399-0144981 |
| 60 | 413.84 | 407.18 | 09:15:55.890 | -50:00:20.00 | 2MASS |
| 61 | 391.00 | 1017.50 | 09:15:33.090 | -50:00:09.80 | 2MASS |
| 62 | 394.39 | 1002.46 | 09:15:33.556 | -50:00:11.64 | 0399-0144780 |
| 63 | 406.50 | 997.00 | 09:15:33.820 | -50:00:15.60 | 2MASS |
| 64 | 409.72 | 554.73 | 09:15:50.413 | -50:00:19.01 | 0399-0145010 |
| 65 | 414.50 | 404.50 | | | |
| 66 | 417.32 | 271.90 | 09:16:01.063 | -50:00:22.29 | 0399-0145189 |
| 67 | 423.06 | 212.60 | 09:16:03.319 | -50:00:24.69 | 0399-0145201 |
| 68 | 425.16 | 559.75 | 09:15:50.270 | -50:00:23.70 | 2MASS |
| 69 | 427.34 | 456.50 | 09:15:54.046 | -50:00:26.67 | 0399-0145080 |
| 70 | 431.56 | 354.23 | 09:15:57.980 | -50:00:26.80 | 2MASS |
| 71 | 433.91 | 987.61 | 09:15:34.194 | -50:00:25.49 | 0399-0144786 |
| 72 | 436.22 | 569.75 | 09:15:49.973 | -50:00:27.10 | 0399-0145001 |
| 73 | 438.85 | 470.83 | 09:15:53.590 | -50:00:29.00 | 2MASS |
| 74 | 444.60 | 488.95 | 09:15:52.949 | -50:00:32.97 | 0399-0145059 |
| 75 | 452.82 | 620.06 | 09:15:47.990 | -50:00:33.80 | 0399-0144977 |
| 76 | 457.23 | 443.99 | 09:15:54.493 | -50:00:35.49 | 0399-0145090 |
| 77 | 460.30 | 167.34 | 09:16:04.968 | -50:00:38.17 | 0399-0145217 |
| 78 | 460.54 | 400.09 | 09:15:56.246 | -50:00:37.44 | 0399-0145114 |
| 79 | 467.10 | 578.76 | 09:15:49.608 | -50:00:39.48 | 0399-0144995 |
| 80 | 474.94 | 818.00 | 09:15:40.547 | -50:00:41.23 | 0399-0144869 |
| 81 | 497.14 | 538.62 | 09:15:51.077 | -50:00:49.61 | 0399-0145019 |
| 82 | 498.50 | 531.50 | 09:15:51.050 | -50:00:49.90 | 2MASS |
| 83 | 499.24 | 931.40 | 09:15:36.318 | -50:00:49.69 | 0399-0144816 |
| 84 | 499.61 | 451.51 | 09:15:54.249 | -50:00:50.99 | 0399-0145086 |
| 85 | 519.13 | 858.22 | 09:15:38.956 | -50:00:56.67 | 0399-0144847 |
| 86 | 478.50 | 492.50 | 09:15:52.787 | -50:00:43.83 | 0399-0145056 |
| 87 | 472.50 | 1021.50 | 09:15:32.929 | -50:00:39.85 | 0399-0144773 |
| 88 | 500.82 | 910.97 | 09:15:36.910 | -50:00:51.47 | 0399-0144823 |
| 89 | 527.10 | 691.64 | 09:15:45.268 | -50:01:00.38 | 0399-0144932 |
| 90 | 536.04 | 200.56 | 09:16:03.690 | -50:01:05.54 | 0399-0145203 |
| 91 | 536.81 | 352.30 | 09:15:58.010 | -50:01:04.90 | 2MASS |
| 92 | 541.44 | 460.29 | 09:15:53.950 | -50:01:06.20 | 2MASS |
| 93 | 537.00 | 534.11 | 09:15:51.195 | -50:01:04.55 | 0399-0145021 |
| 94 | 545.66 | 473.02 | 09:15:53.500 | -50:01:08.51 | 0399-0145072 |
| 95 | 548.20 | 348.39 | 09:15:58.150 | -50:01:09.10 | 2MASS |
| 96 | 563.32 | 421.27 | 09:15:55.422 | -50:01:14.53 | 0399-0145103 |
| 97 | 547.72 | 878.77 | 09:15:38.253 | -50:01:07.43 | 0399-0144837 |
| 98 | 572.38 | 170.30 | 09:16:04.826 | -50:01:18.66 | 0399-0145216 |
| 99 | 588.10 | 394.28 | 09:15:56.401 | -50:01:23.89 | 0399-0145116 |
| 100 | 588.81 | 654.51 | 09:15:46.634 | -50:01:23.10 | 0399-0144957 |
| 101 | 591.63 | 814.55 | 09:15:40.615 | -50:01:23.54 | 0399-0144871 |
| 102 | 602.80 | 447.93 | 09:15:54.446 | -50:01:28.88 | 0399-0145089 |
| 103 | 601.58 | 438.82 | 09:15:55.930 | -50:01:31.60 | 2MASS |
| 104 | 610.71 | 407.23 | 09:15:55.881 | -50:01:31.64 | 0399-0145109 |
| 105 | 627.90 | 836.04 | 09:15:39.807 | -50:01:36.49 | 0399-0144856 |
| 106 | 638.48 | 517.76 | 09:15:51.763 | -50:01:41.63 | 0399-0145030 |
| 107 | 641.98 | 453.35 | 09:15:54.167 | -50:01:42.71 | 0399-0145082 |
| 108 | 655.03 | 385.01 | 09:15:56.754 | -50:01:48.15 | 0399-0145124 |
| 109 | 663.20 | 518.02 | 09:15:51.781 | -50:01:50.00 | 0399-0145032 |
| 110 | 697.75 | 302.72 | 09:15:59.832 | -50:02:03.71 | 0399-0145174 |
| 111 | 704.99 | 670.22 | 09:15:46.109 | -50:02:05.21 | 0399-0144941 |
| 112 | 707.83 | 655.52 | 09:15:46.421 | -50:02:06.43 | 0399-0144953 |
| 113 | 724.10 | 938.98 | 09:15:35.952 | -50:02:10.76 | 0399-0144807 |
| 114 | 510.50 | 1.50 | 09:16:11.254 | -50:00:56.74 | 0399-0145276 |
| 115 | 618.81 | 82.42 | 09:16:08.104 | -50:01:35.77 | 0399-0145242 |
| 116 | 743.71 | 50.69 | 09:16:09.166 | -50:02:21.76 | 0399-0145258 |
| 117 | 744.89 | 961.76 | 09:15:35.034 | -50:02:18.38 | 0399-0144792 |

**Table 2.** $(X, Y)$ position on the map of stars with photometry in the field. USNO-B1 identification for these stars and their coordinates.

| Number | X (Pixels) | Y (Pixels) | RA (J2000) | DEC (J2000) | Name (USNO-B1.0) |
|---|---|---|---|---|---|
| 118 | 750.69 | 187.55 | 09:16:04.133 | -50:02:23.27 | 0399-0145210 |
| 119 | 756.20 | 110.99 | 09:16:06.940 | -50:02:25.49 | 0399-0145233 |
| 120 | 774.74 | 386.25 | 09:15:56.639 | -50:02:31.35 | 0399-0145122 |
| 121 | 776.73 | 133.50 | 09:16:06.155 | -50:02:32.37 | 0399-0145220 |
| 122 | 786.70 | 895.70 | 09:15:37.505 | -50:02:34.08 | 0399-0144831 |
| 123 | 796.37 | 508.77 | 09:15:52.053 | -50:02:38.93 | 0399-0145040 |
| 124 | 807.60 | 690.29 | 09:15:45.196 | -50:02:41.77 | 0399-0144930 |
| 125 | 810.26 | 821.51 | 09:15:40.248 | -50:02:42.28 | 0399-0144863 |
| 126 | 834.34 | 917.53 | 09:15:36.645 | -50:02:50.75 | 0399-0144820 |
| 127 | 862.59 | 817.02 | 09:15:40.337 | -50:03:01.63 | 0399-0144865 |
| 128 | 876.56 | 720.85 | 09:15:44.057 | -50:03:07.02 | 0399-0144916 |
| 129 | 885.45 | 372.34 | 09:15:57.132 | -50:03:11.81 | 0399-0145130 |
| 130 | 886.87 | 762.66 | 09:15:42.471 | -50:03:10.38 | 0399-0144898 |
| 131 | 888.96 | 609.71 | 09:15:48.207 | -50:03:12.29 | 0399-0144980 |
| 132 | 903.35 | 810.82 | 09:15:40.656 | -50:03:16.05 | 0399-0144872 |
| 133 | 911.16 | 96.63 | 09:16:07.488 | -50:03:21.76 | 0399-0145237 |
| 134 | 912.48 | 959.80 | 09:15:35.049 | -50:03:19.20 | 0399-0144793 |
| 135 | 914.42 | 219.00 | 09:16:02.893 | -50:03:23.26 | 0399-0145197 |
| 136 | 924.59 | 508.77 | 09:15:51.975 | -50:03:25.47 | 0399-0145037 |
| 137 | 939.09 | 63.46 | 09:16:08.695 | -50:03:31.74 | 0399-0145247 |
| 138 | 948.55 | 511.12 | 09:15:51.878 | -50:03:33.39 | 0399-0145034 |
| 139 | 949.80 | 985.86 | 09:15:34.047 | -50:03:32.64 | 0399-0144785 |
| 140 | 950.30 | 539.49 | 09:15:50.899 | -50:03:34.22 | 0399-0145017 |
| 141 | 963.85 | 648.05 | 09:15:46.743 | -50:03:38.76 | 0399-0144960 |
| 142 | 980.46 | 832.07 | 09:15:39.837 | -50:03:44.11 | 0399-0144857 |
| 143 | 1005.65 | 256.70 | 09:16:01.446 | -50:03:55.25 | 0399-0145191 |
| 144 | 1013.03 | 786.99 | 09:15:41.510 | -50:03:56.42 | 0399-0144881 |
| 145 | 1012.74 | 122.30 | 09:16:06.483 | -50:03:58.28 | 0399-0145227 |
| 146 | 1016.38 | 409.28 | 09:15:55.692 | -50:03:58.87 | 0399-0145105 |
| 147 | 668.20 | 1017.34 | 09:15:32.978 | -50:01:50.30 | 0399-0144774 |

**Table 3.** Photometry for stars in Pismis 11.

| N⁰ | V | $\sigma_V$ | N | (V − R) | $\sigma_{(V-R)}$ | N | (V − I) | $\sigma_{(V-I)}$ | N | (U − B) | $\sigma_{(U-B)}$ | N | B | $\sigma_B$ | N |
|---|---|---|---|---|---|---|---|---|---|---|---|---|---|---|---|
| 1 | 15.722 | 0.031 | 5 | 0.591 | 0.040 | 5 | 1.282 | 0.029 | 2 | | | | | | |
| 2 | 15.358 | 0.026 | 7 | 0.671 | 0.042 | 7 | 1.493 | 0.040 | 2 | 0.401 | 0.056 | 7 | 16.402 | 0.053 | 7 |
| 3 | 14.464 | 0.030 | 5 | 0.423 | 0.036 | 5 | 0.874 | 0.000 | 1 | 0.303 | 0.000 | 1 | 15.238 | 0.000 | 1 |
| 4 | 16.814 | 0.030 | 7 | 0.784 | 0.065 | 7 | 1.563 | 0.040 | 2 | - | - | - | - | - | - |
| 5 | 14.568 | 0.026 | 4 | 0.678 | 0.019 | 4 | 1.360 | 0.000 | 1 | -0.024 | 0.069 | 3 | 15.554 | 0.090 | 3 |
| 6 | 17.362 | 0.026 | 7 | 0.722 | 0.039 | 7 | 1.511 | 0.047 | 2 | - | - | - | - | - | - |
| 7 | - | - | - | - | - | - | - | - | - | 0.441 | 0.065 | 3 | 16.403 | 0.015 | 3 |
| 8 | 15.615 | 0.032 | 6 | 1.553 | 0.056 | 6 | 2.953 | 0.047 | 2 | 2.308 | 0.000 | 1 | 18.166 | 0.000 | 1 |
| 9 | 12.646 | 0.042 | 3 | 0.572 | 0.059 | 3 | 1.219 | 0.000 | 1 | 0.745 | 0.026 | 2 | 13.522 | 0.034 | 2 |
| 10 | 17.811 | 0.042 | 5 | 0.925 | 0.027 | 5 | 1.793 | 0.006 | 2 | 0.620 | 0.167 | 3 | 19.232 | 0.067 | 3 |
| 11 | 13.196 | 0.000 | 1 | 0.310 | 0.000 | 1 | 0.782 | 0.000 | 1 | 0.520 | 0.053 | 6 | 13.932 | 0.059 | 6 |
| 12 | 17.422 | 0.032 | 4 | 0.935 | 0.051 | 4 | 1.822 | 0.000 | 1 | 0.685 | 0.088 | 5 | 18.874 | 0.075 | 5 |
| 13 | - | - | - | - | - | - | - | - | - | 0.469 | 0.059 | 3 | 16.550 | 0.023 | 3 |
| 14 | 16.307 | 0.024 | 7 | 1.016 | 0.053 | 7 | 2.030 | 0.045 | 2 | 0.630 | 0.035 | 3 | 17.648 | 0.042 | 3 |
| 15 | 15.199 | 0.030 | 6 | 0.613 | 0.044 | 6 | 1.352 | 0.032 | 2 | 0.358 | 0.042 | 5 | 16.247 | 0.058 | 5 |
| 16 | 14.340 | 0.025 | 6 | 0.417 | 0.047 | 6 | 0.915 | 0.023 | 2 | 0.217 | 0.046 | 6 | 15.081 | 0.042 | 6 |
| 17 | 12.631 | 0.040 | 2 | 0.330 | 0.063 | 2 | | | 0 | -0.019 | 0.008 | 3 | 13.260 | 0.031 | 3 |
| 18 | 14.486 | 0.036 | 6 | 0.476 | 0.039 | 6 | 1.041 | 0.040 | 2 | 0.367 | 0.034 | 5 | 15.342 | 0.043 | 5 |
| 19 | 15.786 | 0.049 | 4 | 0.724 | 0.064 | 4 | 1.478 | 0.035 | 2 | 0.386 | 0.044 | 6 | 16.859 | 0.048 | 6 |
| 20 | 14.423 | 0.023 | 6 | 0.470 | 0.035 | 6 | 0.924 | 0.066 | 2 | 0.176 | 0.060 | 4 | 15.088 | 0.046 | 4 |
| 21 | 17.539 | 0.032 | 7 | 0.771 | 0.048 | 7 | 1.501 | 0.054 | 2 | 0.559 | 0.092 | 6 | 18.787 | 0.065 | 6 |
| 22 | 16.488 | 0.044 | 7 | 1.564 | 0.062 | 7 | 2.979 | 0.045 | 2 | 1.865 | 0.000 | 1 | 19.061 | 0.000 | 1 |
| 23 | 15.666 | 0.026 | 5 | 0.730 | 0.045 | 5 | 1.600 | 0.035 | 2 | 0.347 | 0.034 | 5 | 16.753 | 0.043 | 5 |
| 24 | 18.100 | 0.030 | 6 | 0.885 | 0.031 | 6 | 1.776 | 0.000 | 1 | 0.566 | 0.014 | 2 | 19.415 | 0.040 | 2 |
| 25 | 18.033 | 0.027 | 6 | 0.922 | 0.039 | 6 | 1.859 | 0.000 | 1 | 1.343 | - | 1 | 19.419 | - | 1 |
| 26 | 16.631 | 0.028 | 6 | 0.586 | 0.043 | 6 | 1.278 | 0.031 | 2 | 0.237 | 0.038 | 4 | 17.622 | 0.053 | 4 |
| 27 | 14.971 | 0.027 | 6 | 0.942 | 0.037 | 6 | 1.900 | 0.052 | 2 | 0.896 | 0.054 | 5 | 16.295 | 0.050 | 5 |
| 28 | 13.451 | 0.043 | 3 | 1.344 | 0.073 | 3 | 2.590 | 0.000 | 1 | 2.260 | 0.026 | 4 | 15.603 | 0.063 | 4 |
| 29 | 16.911 | 0.019 | 2 | 0.751 | 0.089 | 2 | - | - | 0 | 0.859 | 0.057 | 2 | 18.142 | 0.023 | 2 |
| 30 | 17.734 | 0.018 | 6 | 0.811 | 0.038 | 6 | 1.576 | 0.047 | 2 | 0.516 | 0.013 | 4 | 18.978 | 0.065 | 4 |
| 31 | 15.011 | 0.024 | 6 | 0.486 | 0.033 | 6 | 1.018 | 0.033 | 2 | 0.513 | 0.031 | 5 | 15.860 | 0.036 | 5 |
| 32 | 15.968 | 0.024 | 6 | 0.874 | 0.033 | 6 | 1.832 | 0.000 | 2 | 1.048 | 0.027 | 5 | 17.248 | 0.044 | 5 |
| 33 | 16.051 | 0.024 | 6 | 0.597 | 0.059 | 6 | 1.190 | 0.000 | 1 | 0.736 | 0.056 | 4 | 16.984 | 0.054 | 4 |
| 35 | 17.341 | 0.028 | 6 | 0.973 | 0.040 | 6 | 2.033 | 0.000 | 1 | 1.116 | 0.086 | 3 | 18.723 | 0.076 | 3 |
| 36 | 17.370 | 0.028 | 7 | 0.757 | 0.036 | 7 | 1.557 | 0.000 | 2 | 0.463 | 0.028 | 3 | 18.572 | 0.026 | 3 |
| 38 | 15.917 | 0.028 | 7 | 0.898 | 0.039 | 7 | 1.882 | 0.017 | 2 | 0.714 | 0.031 | 5 | 17.226 | 0.047 | 5 |
| 39 | 16.452 | 0.022 | 6 | 0.709 | 0.035 | 6 | 1.547 | 0.025 | 2 | 0.599 | 0.061 | 6 | 17.515 | 0.051 | 6 |
| 40 | 16.784 | 0.025 | 6 | 0.743 | 0.038 | 6 | 1.617 | 0.020 | 2 | 0.737 | 0.033 | 5 | 17.897 | 0.052 | 5 |
| 41 | 15.988 | 0.023 | 6 | 0.828 | 0.036 | 6 | 1.789 | 0.037 | 2 | 0.664 | 0.035 | 5 | 17.202 | 0.041 | 5 |
| 42 | 15.951 | 0.029 | 7 | 0.700 | 0.044 | 7 | 1.500 | 0.024 | 2 | 0.447 | 0.035 | 5 | 16.969 | 0.044 | 5 |
| 43 | 17.514 | 0.035 | 7 | 0.951 | 0.047 | 7 | 1.879 | 0.022 | 2 | 0.705 | 0.086 | 4 | 18.946 | 0.065 | 4 |
| 44 | 17.481 | 0.030 | 7 | 0.867 | 0.042 | 7 | 1.799 | 0.020 | 2 | 1.037 | 0.061 | 4 | 18.721 | 0.071 | 4 |
| 45 | 17.381 | 0.028 | 7 | 0.972 | 0.037 | 7 | 1.949 | 0.001 | 2 | 0.873 | 0.015 | 3 | 18.884 | 0.072 | 3 |

**Table 3.** Photometry for stars in Pismis 11.

| N° | V | $\sigma_V$ | N | $(V-R)$ | $\sigma_{(V-R)}$ | N | $(V-I)$ | $\sigma_{(V-I)}$ | N | $(U-B)$ | $\sigma_{(U-B)}$ | N | B | $\sigma_B$ | N |
|---|---|---|---|---|---|---|---|---|---|---|---|---|---|---|---|
| 46 | 18.155 | 0.030 | 6 | 0.854 | 0.048 | 6 | 1.713 | 0.008 | 2 | 0.630 | 0.077 | 3 | 19.491 | 0.039 | 3 |
| 47 | 16.424 | 0.022 | 6 | 0.913 | 0.031 | 6 | 1.942 | 0.049 | 2 | 1.016 | 0.032 | 5 | 17.764 | 0.048 | 5 |
| 48 | 18.029 | 0.029 | 6 | 0.856 | 0.032 | 6 | 1.788 | 0.011 | 2 | 0.680 | 0.083 | 3 | 19.306 | 0.051 | 3 |
| 49 | 12.898 | 0.036 | 3 | 0.656 | 0.060 | 3 | 1.336 | 0.000 | 1 | -0.177 | 0.050 | 4 | 13.854 | 0.059 | 4 |
| 50 | 16.473 | 0.032 | 7 | 0.751 | 0.039 | 7 | 1.592 | 0.000 | 1 | 0.538 | 0.051 | 6 | 17.572 | 0.052 | 6 |
| 51 | 16.860 | 0.028 | 7 | 1.016 | 0.044 | 7 | 2.068 | 0.000 | 2 | 0.950 | 0.034 | 5 | 18.379 | 0.049 | 5 |
| 52 | 17.260 | 0.029 | 6 | 0.736 | 0.041 | 6 | 1.545 | 0.041 | 2 | 0.656 | 0.028 | 4 | 18.539 | 0.064 | 4 |
| 53 | 16.568 | 0.031 | 7 | 0.763 | 0.043 | 7 | 1.609 | 0.000 | 1 | 0.614 | 0.046 | 6 | 17.671 | 0.053 | 6 |
| 54 | 16.801 | 0.031 | 7 | 0.724 | 0.043 | 7 | 1.552 | 0.045 | 2 | 0.679 | 0.025 | 5 | 17.846 | 0.047 | 5 |
| 55 | 17.985 | 0.031 | 6 | 0.887 | 0.038 | 6 | 1.758 | 0.033 | 2 | 1.300 | 0.000 | 1 | 19.454 | 0.000 | 1 |
| 56 | 15.561 | 0.025 | 6 | 0.720 | 0.043 | 6 | 1.566 | 0.008 | 2 | 0.642 | 0.023 | 5 | 16.615 | 0.051 | 5 |
| 57 | 16.641 | 0.026 | 6 | 0.724 | 0.033 | 6 | 1.584 | 0.031 | 2 | 0.677 | 0.052 | 5 | 17.723 | 0.056 | 5 |
| 58 | 17.540 | 0.030 | 6 | 0.799 | 0.029 | 6 | 1.730 | 0.034 | 2 | 1.024 | 0.067 | 4 | 18.749 | 0.059 | 4 |
| 59 | 14.258 | 0.013 | 4 | 0.755 | 0.036 | 4 | - | - | 0 | 0.046 | 0.046 | 6 | 15.445 | 0.038 | 6 |
| 60 | 15.567 | 0.031 | 7 | 0.724 | 0.044 | 7 | 1.541 | 0.010 | 2 | 0.857 | 0.014 | 4 | 16.670 | 0.048 | 4 |
| 61 | 17.141 | 0.022 | 4 | 0.673 | 0.045 | 4 | 1.324 | 0.000 | 1 | 0.445 | 0.097 | 2 | 18.261 | 0.088 | 2 |
| 62 | 14.371 | 0.026 | 4 | 0.547 | 0.044 | 4 | 1.111 | 0.000 | 1 | 0.149 | 0.086 | 2 | 15.210 | 0.080 | 2 |
| 63 | 17.675 | 0.019 | 5 | 0.719 | 0.046 | 5 | 1.413 | 0.037 | 2 | 0.749 | 0.069 | 3 | 18.876 | 0.030 | 3 |
| 64 | 18.453 | 0.032 | 6 | 0.931 | 0.033 | 6 | 1.941 | 0.000 | 1 | 0.669 | 0.000 | 1 | 19.921 | 0.000 | 1 |
| 65 | 15.951 | 0.029 | 7 | 0.647 | 0.044 | 7 | 1.409 | 0.030 | 2 | 0.560 | 0.057 | 5 | 16.832 | 0.065 | 5 |
| 66 | 16.755 | 0.025 | 6 | 0.725 | 0.069 | 6 | 1.542 | 0.027 | 2 | 0.772 | 0.050 | 5 | 17.740 | 0.050 | 5 |
| 67 | 19.819 | 0.040 | 4 | -0.106 | 0.063 | 4 | -0.087 | 0.000 | 1 | -0.555 | 0.065 | 4 | 19.920 | 0.077 | 4 |
| 68 | 16.996 | 0.022 | 6 | 0.893 | 0.039 | 6 | 1.875 | 0.000 | 1 | 1.023 | 0.048 | 6 | 18.311 | 0.053 | 6 |
| 69 | 12.681 | 0.034 | 3 | 0.638 | 0.060 | 3 | 1.317 | 0.000 | 1 | -0.161 | 0.051 | 4 | 13.622 | 0.059 | 4 |
| 70 | 15.338 | 0.029 | 7 | 0.625 | 0.040 | 7 | 1.363 | 0.041 | 2 | 0.245 | 0.030 | 5 | 16.266 | 0.039 | 5 |
| 71 | 15.035 | 0.036 | 6 | 0.486 | 0.045 | 6 | 0.897 | 0.000 | 1 | 0.347 | 0.064 | 4 | 15.830 | 0.049 | 4 |
| 72 | 13.533 | 0.033 | 3 | 0.720 | 0.054 | 3 | 1.489 | 0.000 | 1 | -0.048 | 0.053 | 4 | 14.604 | 0.061 | 4 |
| 73 | 16.800 | 0.033 | 6 | 0.852 | 0.035 | 6 | 1.749 | 0.000 | 1 | 0.946 | 0.024 | 5 | 18.011 | 0.054 | 5 |
| 74 | 16.562 | 0.021 | 8 | 0.739 | 0.040 | 8 | 1.551 | 0.004 | 2 | 0.739 | 0.022 | 5 | 17.618 | 0.051 | 5 |
| 75 | 16.283 | 0.012 | 4 | 0.583 | 0.039 | 4 | - | - | 0 | 0.453 | 0.034 | 5 | 17.313 | 0.049 | 5 |
| 76 | 17.043 | 0.025 | 8 | 1.125 | 0.037 | 8 | 2.242 | 0.004 | 2 | 0.963 | 0.046 | 4 | 18.769 | 0.073 | 4 |
| 77 | 16.080 | 0.032 | 7 | 0.802 | 0.045 | 7 | 1.672 | 0.029 | 2 | 0.912 | 0.032 | 4 | 17.328 | 0.061 | 4 |
| 78 | 16.037 | 0.032 | 8 | 0.711 | 0.044 | 8 | 1.497 | 0.013 | 2 | 0.451 | 0.038 | 5 | 17.051 | 0.042 | 5 |
| 79 | 15.409 | 0.029 | 8 | 0.551 | 0.041 | 8 | 1.159 | 0.004 | 2 | 0.186 | 0.034 | 7 | 16.294 | 0.064 | 7 |
| 80 | 16.459 | 0.028 | 6 | 0.590 | 0.071 | 6 | 1.125 | 0.000 | 1 | 0.233 | 0.061 | 6 | 17.446 | 0.050 | 6 |
| 81 | 15.833 | 0.027 | 7 | 0.702 | 0.032 | 7 | 1.512 | 0.004 | 2 | 0.604 | 0.043 | 7 | 16.882 | 0.064 | 7 |
| 82 | 17.636 | 0.036 | 6 | 0.938 | 0.037 | 6 | 1.978 | 0.001 | 2 | 1.036 | 0.077 | 3 | 19.009 | 0.129 | 3 |
| 83 | 14.750 | 0.036 | 6 | 0.503 | 0.041 | 6 | 1.031 | 0.000 | 1 | 0.546 | 0.049 | 6 | 15.479 | 0.045 | 6 |
| 84 | 16.793 | 0.031 | 8 | 0.751 | 0.045 | 8 | 1.596 | 0.021 | 2 | 0.898 | 0.023 | 5 | 17.875 | 0.053 | 5 |
| 85 | 16.894 | 0.029 | 6 | 0.856 | 0.048 | 6 | 1.656 | 0.000 | 1 | 0.591 | 0.057 | 6 | 18.275 | 0.057 | 5 |
| 86 | - | - | - | - | - | - | - | - | - | -0.311 | 0.035 | 3 | 12.122 | 0.032 | 3 |
| 87 | - | - | - | - | - | - | - | - | - | -0.043 | 0.000 | 1 | 12.669 | 0.000 | 1 |
| 88 | 17.813 | 0.036 | 4 | 0.978 | 0.023 | 4 | 1.945 | 0.010 | 2 | 0.900 | 0.033 | 2 | 19.183 | 0.045 | 2 |

**Table 3.** Photometry for stars in Pismis 11.

| $N^0$ | V | $\sigma_V$ | N | $(V-R)$ | $\sigma_{(V-R)}$ | N | $(V-I)$ | $\sigma_{(V-I)}$ | N | $(U-B)$ | $\sigma_{(U-B)}$ | N | B | $\sigma_B$ | N |
|---|---|---|---|---|---|---|---|---|---|---|---|---|---|---|---|
| 89 | 15.388 | 0.038 | 5 | 1.098 | 0.014 | 5 | 2.113 | 0.000 | 1 | 1.128 | 0.051 | 6 | 16.936 | 0.059 | 6 |
| 90 | 17.741 | 0.022 | 5 | 0.780 | 0.034 | 5 | 1.598 | 0.000 | 2 | 0.552 | 0.069 | 4 | 19.051 | 0.099 | 4 |
| 91 | 16.366 | 0.028 | 8 | 0.673 | 0.045 | 8 | 1.386 | 0.011 | 2 | 0.436 | 0.024 | 5 | 17.411 | 0.042 | 5 |
| 92 | 16.491 | 0.027 | 8 | 0.736 | 0.036 | 8 | 1.578 | 0.021 | 2 | 0.651 | 0.032 | 5 | 17.567 | 0.048 | 5 |
| 93 | 18.135 | 0.028 | 7 | 0.898 | 0.052 | 7 | 1.889 | 0.016 | 2 | 0.598 | 0.033 | 2 | 19.481 | 0.049 | 2 |
| 94 | 14.633 | 0.030 | 8 | 0.698 | 0.042 | 8 | 1.480 | 0.013 | 2 | 0.219 | 0.036 | 7 | 15.692 | 0.057 | 7 |
| 95 | 15.945 | 0.021 | 6 | 0.702 | 0.037 | 6 | 1.515 | 0.030 | 2 | 0.260 | 0.050 | 6 | 16.939 | 0.046 | 6 |
| 96 | 16.471 | 0.029 | 7 | 0.578 | 0.035 | 7 | 1.215 | 0.020 | 2 | 0.415 | 0.026 | 5 | 17.432 | 0.041 | 5 |
| 97 | 17.993 | 0.032 | 5 | 0.889 | 0.036 | 5 | 1.774 | 0.000 | 1 | 0.545 | 0.053 | 2 | 19.326 | 0.005 | 2 |
| 98 | 16.618 | 0.026 | 6 | 0.862 | 0.036 | 6 | 1.719 | 0.025 | 2 | 1.273 | 0.039 | 4 | 18.000 | 0.063 | 4 |
| 99 | 15.539 | 0.024 | 6 | 0.487 | 0.032 | 6 | 1.065 | 0.016 | 2 | 0.195 | 0.031 | 5 | 16.323 | 0.040 | 5 |
| 100 | 16.237 | 0.015 | 3 | 0.707 | 0.054 | 3 | 1.438 | 0.022 | 2 | 0.566 | 0.037 | 5 | 17.212 | 0.046 | 5 |
| 101 | 16.769 | 0.040 | 6 | 0.694 | 0.045 | 6 | 1.418 | 0.000 | 1 | 0.739 | 0.061 | 6 | 17.798 | 0.057 | 6 |
| 102 | 16.142 | 0.023 | 6 | 0.685 | 0.036 | 6 | 1.501 | 0.009 | 2 | 0.514 | 0.033 | 5 | 17.152 | 0.051 | 5 |
| 103 | 17.996 | 0.029 | 7 | 0.837 | 0.036 | 7 | 1.715 | 0.069 | 2 | 0.938 | 0.128 | 2 | 19.145 | 0.076 | 2 |
| 104 | 16.385 | 0.031 | 7 | 0.617 | 0.042 | 7 | 1.303 | 0.032 | 2 | 0.406 | 0.025 | 5 | 17.386 | 0.042 | 5 |
| 105 | 18.003 | 0.045 | 6 | 0.984 | 0.043 | 6 | 1.932 | 0.000 | 1 | 0.712 | 0.052 | 3 | 19.515 | 0.091 | 3 |
| 106 | 13.360 | 0.040 | 3 | 0.585 | 0.068 | 3 | 1.214 | 0.000 | 1 | -0.164 | 0.030 | 3 | 14.201 | 0.053 | 3 |
| 107 | 12.159 | 0.038 | 3 | 0.467 | 0.057 | 3 | 0.904 | 0.000 | 1 | 0.307 | 0.025 | 2 | 13.001 | 0.030 | 2 |
| 108 | 17.441 | 0.022 | 7 | 1.008 | 0.028 | 7 | 2.089 | 0.024 | 2 | 0.998 | 0.035 | 4 | 18.844 | 0.057 | 4 |
| 109 | 18.316 | 0.029 | 7 | 0.939 | 0.044 | 7 | 1.882 | 0.004 | 2 | 0.770 | 0.004 | 2 | 19.692 | 0.049 | 2 |
| 110 | 17.152 | 0.027 | 7 | 0.940 | 0.038 | 7 | 1.938 | 0.040 | 2 | 0.806 | 0.094 | 4 | 18.632 | 0.082 | 4 |
| 111 | 16.190 | 0.016 | 3 | 0.558 | 0.067 | 3 | 1.102 | 0.031 | 2 | 0.226 | 0.027 | 5 | 17.044 | 0.046 | 5 |
| 112 | 16.480 | 0.011 | 3 | 0.688 | 0.053 | 3 | 1.347 | 0.023 | 2 | 0.494 | 0.025 | 5 | 17.539 | 0.047 | 5 |
| 113 | 17.496 | 0.036 | 7 | 0.664 | 0.063 | 7 | 1.278 | 0.065 | 2 | 0.382 | 0.048 | 6 | 18.595 | 0.039 | 6 |
| 114 | 18.047 | 0.029 | 6 | 0.927 | 0.040 | 6 | 1.899 | 0.033 | 2 | 0.853 | - | 1 | 19.458 | - | 1 |
| 115 | 18.380 | 0.027 | 6 | 0.945 | 0.038 | 6 | 1.932 | 0.000 | 1 | 0.811 | - | 1 | 19.844 | - | 1 |
| 116 | 17.437 | 0.026 | 7 | 0.831 | 0.046 | 7 | 1.696 | 0.039 | 2 | 0.614 | 0.044 | 3 | 18.735 | 0.066 | 3 |
| 117 | 16.972 | 0.035 | 6 | 0.763 | 0.044 | 6 | 1.474 | 0.000 | 1 | 0.491 | 0.070 | 5 | 18.152 | 0.055 | 5 |
| 118 | 17.534 | 0.027 | 7 | 0.867 | 0.060 | 7 | 1.823 | 0.045 | 2 | 0.791 | 0.026 | 3 | 18.881 | 0.059 | 3 |
| 119 | 14.299 | 0.024 | 7 | 0.535 | 0.047 | 7 | 1.145 | 0.042 | 2 | 0.257 | 0.033 | 4 | 15.112 | 0.051 | 4 |
| 120 | 14.848 | 0.029 | 7 | 0.780 | 0.037 | 7 | 1.637 | 0.025 | 2 | 0.618 | 0.038 | 6 | 16.073 | 0.045 | 6 |
| 121 | 17.459 | 0.022 | 7 | 0.705 | 0.046 | 7 | 1.440 | 0.036 | 2 | 0.782 | 0.052 | 4 | 18.642 | 0.066 | 4 |
| 122 | 13.614 | 0.028 | 4 | 0.367 | 0.015 | 4 | 0.700 | 0.000 | 1 | 0.410 | 0.056 | 4 | 14.117 | 0.058 | 4 |
| 123 | 12.952 | 0.035 | 3 | 0.627 | 0.072 | 3 | 1.309 | 0.000 | 1 | -0.107 | 0.028 | 3 | 13.865 | 0.058 | 3 |
| 124 | - | - | - | - | - | - | - | - | - | 1.436 | 0.056 | 4 | 13.461 | 0.065 | 4 |
| 125 | 16.083 | 0.031 | 5 | 0.563 | 0.055 | 5 | 1.115 | 0.000 | 1 | 0.178 | 0.043 | 6 | 16.973 | 0.048 | 6 |
| 126 | 18.176 | 0.036 | 6 | 0.920 | 0.040 | 6 | 1.797 | 0.000 | 1 | 0.577 | - | 1 | 19.589 | - | 1 |
| 127 | 15.438 | 0.034 | 5 | 0.560 | 0.056 | 5 | 1.118 | 0.000 | 1 | 0.242 | 0.036 | 5 | 16.303 | 0.042 | 5 |
| 128 | 15.376 | 0.033 | 5 | 0.543 | 0.082 | 5 | 1.026 | 0.000 | 1 | 0.319 | 0.031 | 5 | 16.308 | 0.039 | 5 |
| 129 | 17.685 | 0.020 | 5 | 0.840 | 0.041 | 5 | 1.631 | 0.027 | 2 | 1.301 | 0.033 | 2 | 19.062 | 0.047 | 2 |
| 130 | 17.960 | 0.035 | 6 | 0.980 | 0.044 | 6 | 1.878 | 0.023 | 2 | 0.654 | - | 1 | 19.500 | - | 1 |
| 131 | - | - | - | - | - | - | - | - | - | 0.714 | 0.034 | 5 | 15.485 | 0.043 | 5 |

**Table 3.** Photometry for stars in Pismis 11.

| N⁰ | V | $\sigma_V$ | N | (V−R) | $\sigma_{(V-R)}$ | N | (V−I) | $\sigma_{(V-I)}$ | N | (U−B) | $\sigma_{(U-B)}$ | N | B | $\sigma_B$ | N |
|---|---|---|---|---|---|---|---|---|---|---|---|---|---|---|---|
| 132 | 17.284 | 0.040 | 7 | 1.180 | 0.073 | 7 | 2.332 | 0.071 | 2 | 0.993 | 0.044 | 2 | 18.997 | 0.022 | 2 |
| 133 | 16.959 | 0.027 | 7 | 0.912 | 0.049 | 7 | 1.895 | 0.000 | 2 | 0.651 | 0.066 | 4 | 18.307 | 0.044 | 4 |
| 134 | 17.941 | 0.031 | 7 | 0.889 | 0.045 | 7 | 1.744 | 0.032 | 2 | 0.594 | 0.023 | 2 | 19.333 | 0.013 | 2 |
| 135 | 15.636 | 0.028 | 7 | 0.620 | 0.040 | 7 | 1.309 | 0.039 | 2 | 0.386 | 0.028 | 4 | 16.645 | 0.047 | 4 |
| 136 | 16.953 | 0.022 | 6 | 0.796 | 0.040 | 6 | 1.685 | 0.041 | 2 | 0.607 | 0.050 | 4 | 18.192 | 0.035 | 4 |
| 137 | 16.048 | 0.028 | 6 | 0.660 | 0.053 | 6 | 1.330 | 0.041 | 2 | 0.521 | 0.045 | 4 | 17.130 | 0.044 | 4 |
| 138 | 13.958 | 0.026 | 4 | 0.530 | 0.037 | 4 | 1.121 | 0.023 | 2 | 0.343 | 0.033 | 5 | 14.837 | 0.041 | 5 |
| 139 | 16.749 | 0.034 | 6 | 0.674 | 0.042 | 6 | 1.302 | 0.000 | 1 | 0.400 | 0.081 | 3 | 17.791 | 0.053 | 3 |
| 140 | 16.706 | 0.028 | 7 | 0.778 | 0.046 | 7 | 1.528 | 0.014 | 2 | 1.141 | 0.024 | 5 | 17.961 | 0.036 | 5 |
| 141 | 17.847 | 0.012 | 3 | 0.920 | 0.037 | 3 | 1.795 | 0.036 | 2 | 0.672 | 0.023 | 2 | 19.240 | 0.013 | 2 |
| 142 | 17.815 | 0.029 | 5 | 0.821 | 0.053 | 5 | 1.619 | 0.025 | 2 | 0.510 | 0.029 | 2 | 19.114 | 0.000 | 2 |
| 143 | 18.197 | 0.027 | 6 | 0.990 | 0.042 | 6 | 2.000 | 0.000 | 2 | 0.881 | 0.000 | 1 | 19.826 | 0.000 | 1 |
| 144 | 15.411 | 0.035 | 5 | 0.566 | 0.060 | 5 | 1.148 | 0.000 | 1 | 0.150 | 0.037 | 5 | 16.303 | 0.034 | 5 |
| 145 | 16.491 | 0.029 | 7 | 0.807 | 0.042 | 7 | 1.715 | 0.043 | 2 | 1.035 | 0.019 | 4 | 17.727 | 0.031 | 4 |
| 146 | 16.267 | 0.027 | 7 | 0.567 | 0.044 | 7 | 1.174 | 0.010 | 2 | 0.223 | 0.040 | 4 | 17.164 | 0.046 | 4 |
| 147 | - | - | - | - | - | - | - | - | - | 0.318 | 0.000 | 1 | 13.323 | 0.000 | 1 |

**Table 5.** Photometry for stars in Alicante 5.

| $N^0$ | $V$ | $\sigma_V$ | $N$ | $(B-V)$ | $\sigma_{(B-V)}$ | $N$ | $(U-B)$ | $\sigma_{(U-B)}$ | $N$ | $(V-R)$ | $\sigma_{(V-R)}$ | $N$ | $(V-I)$ | $\sigma_{(V-I)}$ | $N$ |
|---|---|---|---|---|---|---|---|---|---|---|---|---|---|---|---|
| 1 | 14.810 | 0.006 | 2 | 0.743 | 0.007 | 2 | 0.512 | 0.011 | 2 | 0.513 | 0.007 | 2 | 0.925 | 0.004 | 1 |
| 2 | 16.242 | 0.001 | 2 | 1.252 | 0.001 | 2 | 0.655 | 0.013 | 2 | 0.751 | 0.003 | 2 | 1.497 | 0.000 | 2 |
| 3 | 16.543 | 0.008 | 2 | 1.194 | 0.008 | 2 | 0.506 | 0.016 | 2 | 0.766 | 0.007 | 2 | 1.533 | 0.005 | 2 |
| 4 | 16.850 | 0.017 | 2 | 1.196 | 0.035 | 2 | 0.849 | 0.024 | 2 | 0.819 | 0.019 | 2 | 1.669 | 0.004 | 1 |
| 5 | 15.744 | 0.018 | 2 | 2.027 | 0.018 | 2 | 1.691 | 0.033 | 2 | 1.215 | 0.024 | 2 | 1.864 | 0.005 | 1 |
| 6 | 17.379 | 0.011 | 2 | 1.442 | 0.011 | 2 | 0.815 | 0.016 | 2 | 0.932 | 0.008 | 2 | 1.912 | 0.026 | 2 |
| 7 | 17.003 | 0.005 | 2 | 1.556 | 0.023 | 2 | 1.239 | 0.029 | 2 | 1.037 | 0.002 | 2 | | | |
| 8 | 17.844 | 0.011 | 2 | 1.295 | 0.020 | 2 | 0.689 | 0.019 | 2 | 0.851 | 0.009 | 2 | 1.755 | 0.059 | 2 |
| 9 | 17.694 | 0.011 | 2 | 1.385 | 0.087 | 2 | 0.821 | 0.013 | 2 | 0.903 | 0.007 | 2 | | | |
| 10 | 17.676 | 0.018 | 2 | 1.352 | 0.044 | 2 | 1.171 | 0.044 | 2 | 0.922 | 0.014 | 2 | 1.952 | 0.039 | 2 |
| 11 | 14.625 | 0.006 | 2 | 0.868 | 0.016 | 2 | 0.669 | 0.008 | 2 | 0.596 | 0.005 | 2 | | | |
| 12 | 17.716 | 0.004 | 2 | 1.622 | 0.033 | 2 | 0.876 | 0.088 | 2 | 1.020 | 0.006 | 2 | 1.685 | 0.011 | 1 |
| 13 | 17.383 | 0.001 | 2 | 1.639 | 0.015 | 2 | 1.252 | 0.052 | 2 | 1.055 | 0.001 | 2 | 2.206 | 0.012 | 2 |
| 14 | 17.561 | 0.004 | 2 | 1.369 | 0.040 | 2 | 1.355 | 0.033 | 2 | 0.854 | 0.004 | 2 | | | |
| 15 | 17.970 | 0.004 | 2 | 1.561 | 0.005 | 2 | 0.806 | 0.066 | 2 | 0.985 | 0.000 | 2 | 2.025 | 0.040 | 2 |
| 16 | 18.435 | 0.001 | 2 | 1.357 | 0.035 | 2 | 0.726 | 0.078 | 2 | 0.833 | 0.001 | 2 | 1.576 | 0.010 | 1 |
| 17 | 18.073 | 0.001 | 2 | 1.464 | 0.024 | 2 | 1.221 | 0.024 | 2 | 0.876 | 0.004 | 2 | | | |
| 18 | 18.203 | 0.005 | 2 | 1.612 | 0.052 | 2 | 1.057 | 0.060 | 1 | 1.040 | 0.014 | 2 | 1.924 | 0.037 | 2 |
| 19 | 18.680 | 0.016 | 2 | 1.434 | 0.013 | 2 | 1.004 | 0.071 | 1 | 1.014 | 0.018 | 2 | | | |
| 20 | 18.820 | 0.021 | 2 | 1.484 | 0.048 | 2 | 0.874 | 0.048 | 2 | 0.964 | 0.027 | 2 | 1.953 | 0.035 | 2 |
| 21 | 15.142 | 0.010 | 2 | 0.930 | 0.018 | 2 | 0.430 | 0.012 | 2 | 0.637 | 0.013 | 2 | 0.920 | 0.011 | 1 |
| 22 | 16.654 | 0.001 | 2 | 1.338 | 0.018 | 2 | 1.003 | 0.011 | 1 | 0.845 | 0.008 | 2 | 1.719 | 0.035 | 2 |
| 23 | 17.535 | 0.013 | 2 | 1.423 | 0.012 | 2 | 1.143 | 0.018 | 1 | 0.926 | 0.013 | 2 | 1.929 | 0.008 | |
| 24 | 17.711 | 0.016 | 2 | 1.406 | 0.033 | 2 | 1.223 | 0.033 | 1 | 0.766 | 0.007 | 2 | 1.732 | 0.014 | 1 |
| 25 | 17.935 | 0.013 | 2 | 1.274 | 0.023 | 2 | 0.871 | 0.029 | 1 | 0.810 | 0.024 | 2 | | | |
| 26 | 17.812 | 0.006 | 2 | 1.415 | 0.019 | 2 | 0.733 | 0.029 | 1 | 0.914 | 0.018 | 2 | | | |
| 27 | 17.776 | 0.014 | 2 | 1.506 | 0.036 | 2 | 1.185 | 0.023 | 1 | 1.098 | 0.009 | 2 | 2.566 | 0.006 | 1 |
| 28 | 18.017 | 0.007 | 2 | 1.921 | 0.023 | 2 | 1.447 | 0.037 | 1 | 1.158 | 0.001 | 2 | 2.308 | 0.005 | 1 |
| 29 | 15.327 | 0.011 | 2 | 0.860 | 0.025 | 2 | 0.469 | 0.056 | 2 | 0.524 | 0.040 | 2 | 2.095 | 0.005 | 2 |
| 30 | 18.606 | 0.008 | 2 | 1.430 | 0.001 | 2 | 1.095 | 0.035 | 1 | 0.907 | 0.011 | 2 | 1.826 | 0.037 | 2 |
| 31 | 18.551 | 0.030 | 2 | 1.671 | 0.025 | 2 | | | | 1.091 | 0.041 | 2 | 2.290 | 0.016 | 1 |
| 32 | 18.765 | 0.004 | 2 | 1.520 | 0.042 | 2 | 0.941 | 0.041 | 1 | 0.975 | 0.008 | 2 | 1.577 | 0.015 | 1 |
| 33 | 18.605 | 0.006 | 2 | 1.756 | 0.071 | 2 | 0.863 | 0.037 | 1 | 1.201 | 0.006 | 2 | | | |
| 34 | 18.949 | 0.005 | 2 | 1.507 | 0.040 | 2 | 0.951 | 0.045 | 1 | 1.010 | 0.006 | 2 | 1.735 | 0.012 | 1 |
| 35 | 18.734 | 0.006 | 2 | 1.825 | 0.071 | 2 | 1.398 | 0.061 | 1 | 1.205 | 0.011 | 2 | 2.085 | 0.012 | 1 |
| 36 | 18.868 | 0.021 | 1 | 1.784 | 0.055 | 1 | | | | 1.223 | 0.018 | 1 | | | |
| 37 | 18.651 | 0.009 | 1 | 2.059 | 0.057 | 1 | | | | 1.285 | 0.026 | 2 | 2.663 | 0.045 | 2 |
| 38 | 18.941 | 0.005 | 2 | 1.854 | 0.069 | 2 | 1.560 | 0.087 | 1 | 1.228 | 0.026 | 2 | 2.583 | 0.017 | 1 |
| 39 | 19.223 | 0.009 | 2 | 1.699 | 0.082 | 2 | | | | 1.106 | 0.010 | 2 | | | |
| 40 | 15.384 | 0.021 | 2 | 0.873 | 0.003 | 2 | 0.488 | 0.037 | 2 | 0.534 | 0.042 | 2 | 0.749 | 0.012 | 1 |
| 41 | 19.560 | 0.040 | 2 | 1.513 | 0.071 | 2 | 1.258 | 0.086 | 1 | 1.015 | 0.037 | 2 | 2.084 | 0.013 | 2 |
| 42 | 19.556 | 0.022 | 1 | 2.045 | 0.121 | 1 | | | | 1.388 | 0.044 | 2 | | | |
| 43 | 13.415 | 0.004 | 2 | 1.795 | 0.011 | 2 | 1.904 | 0.011 | 2 | 0.978 | 0.002 | 2 | 1.959 | 0.001 | 1 |
| 44 | 15.429 | 0.011 | 2 | 1.113 | 0.011 | 2 | 0.606 | 0.011 | 2 | 0.744 | 0.006 | 2 | 1.603 | 0.006 | 1 |

**Table 5.** Photometry for stars in Alicante 5.

| $N^0$ | $V$ | $\sigma_V$ | $N$ | $(B-V)$ | $\sigma_{(B-V)}$ | $N$ | $(U-B)$ | $\sigma_{(U-B)}$ | $N$ | $(V-R)$ | $\sigma_{(V-R)}$ | $N$ | $(V-I)$ | $\sigma_{(V-I)}$ | $N$ |
|---|---|---|---|---|---|---|---|---|---|---|---|---|---|---|---|
| 45 | 15.953 | 0.002 | 2 | 1.018 | 0.001 | 2 | 0.349 | 0.019 | 2 | 0.648 | 0.001 | 2 | 1.259 | 0.052 | 2 |
| 46 | 15.834 | 0.025 | 2 | 1.195 | 0.017 | 2 | 1.062 | 0.012 | 2 | 1.014 | 0.033 | 2 | 1.283 | 0.006 | 1 |
| 47 | 13.259 | 0.011 | 2 | 1.204 | 0.018 | 2 | 0.164 | 0.027 | 2 | 0.836 | 0.046 | 2 | | | |
| 48 | 16.312 | 0.013 | 2 | 1.658 | 0.008 | 2 | 1.093 | 0.011 | 1 | 1.108 | 0.065 | 2 | | | |
| 49 | 17.180 | 0.013 | 2 | 1.442 | 0.027 | 2 | 1.099 | 0.011 | 1 | 1.048 | 0.015 | 2 | 2.193 | 0.003 | 1 |
| 50 | 17.346 | 0.012 | 2 | 1.506 | 0.009 | 2 | 1.046 | 0.011 | 1 | 0.937 | 0.029 | 2 | 1.719 | 0.015 | 1 |
| 51 | 17.646 | 0.006 | 1 | 1.671 | 0.020 | 1 | 1.395 | 0.049 | 1 | 1.071 | 0.004 | 2 | 2.179 | 0.009 | 1 |
| 52 | 15.948 | 0.016 | 2 | 2.372 | 0.025 | 2 | 2.594 | 0.011 | 1 | 1.320 | 0.013 | 2 | 2.531 | 0.001 | 1 |
| 53 | 18.060 | 0.001 | 2 | 1.708 | 0.028 | 2 | 1.162 | 0.011 | 1 | 1.053 | 0.007 | 1 | 1.964 | 0.008 | 1 |
| 54 | 18.351 | 0.029 | 2 | 1.504 | 0.064 | 2 | | | | 0.934 | 0.006 | 1 | 1.4 | 0.012 | 1 |
| 55 | 14.419 | 0.012 | 2 | 1.247 | 0.017 | 2 | 0.403 | 0.025 | 2 | 0.836 | 0.037 | 2 | 1.337 | 0.011 | 1 |
| 56 | 17.754 | 0.023 | 2 | 1.559 | 0.035 | 2 | | | | | | 1 | 1.854 | 0.034 | 1 |
| 57 | 18.081 | 0.021 | 2 | 1.621 | 0.038 | 2 | | | | 0.973 | 0.006 | 1 | 1.865 | 0.009 | 1 |
| 58 | 17.905 | 0.031 | 2 | 1.799 | 0.045 | 2 | | | | 1.016 | 0.016 | 1 | | | |
| 59 | 18.163 | 0.020 | 2 | 1.647 | 0.003 | 2 | | | | 0.929 | 0.03 | 1 | 1.999 | 0.039 | 1 |
| 60 | 18.503 | 0.015 | 2 | 1.464 | 0.035 | 2 | | | | 0.799 | 0.008 | 1 | | | |
| 61 | 18.840 | 0.033 | 2 | 1.748 | 0.017 | 2 | | | | 1.081 | 0.002 | 1 | 2.127 | 0.009 | 1 |
| 62 | 15.130 | 0.013 | 2 | 1.222 | 0.019 | 2 | 0.458 | 0.028 | 2 | 0.805 | 0.014 | 2 | 1.566 | 0.006 | 1 |
| 63 | 19.058 | 0.014 | 2 | 1.736 | 0.030 | 2 | | | | 1.003 | 0.056 | 2 | 2.029 | 0.017 | 1 |
| 64 | 18.989 | 0.002 | 1 | 1.741 | 0.027 | 1 | | | | 1.075 | 0.050 | 2 | 2.094 | 0.012 | 1 |
| 65 | 19.026 | 0.011 | 2 | 1.853 | 0.028 | 2 | | | | 1.103 | 0.030 | 2 | 2.144 | 0.011 | 1 |
| 66 | 18.402 | 0.022 | 2 | 2.356 | 0.028 | 2 | | | | 1.425 | 0.048 | 2 | 2.526 | 0.014 | 1 |
| 67 | 14.892 | 0.005 | 2 | 1.505 | 0.011 | 2 | 0.464 | 0.012 | 2 | 1.022 | 0.018 | 2 | 1.976 | 0.001 | 1 |
| 68 | 16.478 | 0.019 | 2 | 1.032 | 0.029 | 2 | 0.544 | 0.011 | 1 | 0.665 | 0.004 | 2 | 1.203 | 0.003 | 1 |
| 69 | 16.275 | 0.012 | 2 | 1.373 | 0.023 | 2 | 0.600 | 0.011 | 1 | 0.895 | 0.010 | 2 | 1.499 | 0.012 | 1 |
| 70 | 16.284 | 0.011 | 2 | 1.373 | 0.023 | 2 | 0.601 | 0.011 | 2 | 0.9 | 0.001 | 2 | 1.718 | 0.002 | 1 |
| 71 | 16.974 | 0.001 | 2 | 1.148 | 0.010 | 2 | 0.553 | 0.011 | 2 | 0.799 | 0.028 | 2 | 1.416 | 0.005 | 1 |
| 72 | 17.691 | 0.004 | 1 | 1.321 | 0.010 | 1 | 1.242 | 0.019 | 1 | 0.849 | 0.012 | 2 | 1.830 | 0.006 | 1 |
| 73 | 18.086 | 0.005 | 1 | 1.567 | 0.014 | 1 | 1.091 | 0.027 | 1 | | | | | | |
| 74 | 18.048 | 0.004 | 1 | 1.475 | 0.022 | 1 | 1.276 | 0.033 | 1 | 0.929 | 0.028 | 2 | 1.688 | 0.015 | 1 |
| 75 | 18.308 | 0.005 | 1 | 1.581 | 0.018 | 1 | 1.006 | 0.033 | 1 | 1.021 | 0.008 | 2 | 1.991 | 0.007 | 1 |
| 76 | 18.363 | 0.006 | 1 | 1.520 | 0.021 | 1 | 1.340 | 0.042 | 1 | 0.922 | 0.041 | | 1.873 | 0.015 | 1 |
| 77 | 18.739 | 0.007 | 1 | 1.542 | 0.023 | 1 | 0.948 | 0.041 | 1 | 1.008 | 0.012 | 2 | 2.087 | 0.011 | 1 |
| 78 | 19.073 | 0.008 | 1 | 1.615 | 0.028 | 1 | 0.856 | 0.055 | 1 | 0.999 | 0.030 | 2 | 2.060 | 0.028 | 2 |
| 79 | 18.764 | 0.008 | 1 | 1.642 | 0.032 | 1 | 1.463 | 0.062 | 1 | 0.955 | 0.026 | 2 | | | |
| 80 | 19.446 | 0.010 | 1 | 1.552 | 0.031 | 1 | 0.811 | 0.058 | 1 | | | | | | |
| 81 | 19.041 | 0.008 | 1 | 1.461 | 0.023 | 1 | 1.399 | 0.056 | 1 | 0.925 | 0.022 | 2 | 1.801 | 0.024 | 2 |
| 82 | 19.114 | 0.009 | 1 | 1.682 | 0.042 | 1 | 1.156 | 0.075 | 1 | 0.979 | 0.054 | 2 | | | |
| 83 | 19.805 | 0.012 | 1 | 1.671 | 0.047 | 1 | 0.920 | 0.108 | 1 | 1.029 | 0.001 | 2 | 1.943 | 0.021 | 1 |
| 84 | 18.530 | 0.006 | 1 | 1.741 | 0.020 | 1 | | | | 1.176 | 0.013 | 2 | 2.379 | 0.010 | 1 |
| 85 | 18.683 | 0.007 | 1 | 1.628 | 0.032 | 1 | | | | 0.989 | 0.054 | 2 | 2.061 | 0.022 | 1 |
| 86 | 19.225 | 0.009 | 1 | 1.646 | 0.031 | 1 | | | | | | | | | |
| 87 | 19.586 | 0.011 | 1 | 1.504 | 0.038 | 1 | | | | 1.054 | 0.037 | 2 | | | |
| 88 | 19.585 | 0.010 | 1 | 1.776 | 0.044 | 1 | | | | 1.223 | 0.033 | | | | |

**Table 5.** Photometry for stars in Alicante 5.

| $N^0$ | $V$ | $\sigma_V$ | $N$ | $(B-V)$ | $\sigma_{(B-V)}$ | $N$ | $(U-B)$ | $\sigma_{(U-B)}$ | $N$ | $(V-R)$ | $\sigma_{(V-R)}$ | $N$ | $(V-I)$ | $\sigma_{(V-I)}$ | $N$ |
|---|---|---|---|---|---|---|---|---|---|---|---|---|---|---|---|
| 89  | 19.998 | 0.014 | 1 | 1.474 | 0.048 | 1 |  |  |  | 0.991 | 0.047 | 2 | 2.105  | 0.026 | 1 |
| 90  | 19.947 | 0.013 | 1 | 1.552 | 0.049 | 1 |  |  |  | 1.058 | 0.062 | 2 | 2.105  | 0.023 | 1 |
| 91  | 20.125 | 0.014 | 1 | 1.553 | 0.054 | 1 |  |  |  |       |       |   |        |       |   |
| 92  | 19.978 | 0.014 | 1 | 1.715 | 0.055 | 1 |  |  |  | 1.062 | 0.012 | 2 |        |       |   |
| 93  | 17.886 | 0.037 | 1 | 1.933 | 0.039 | 1 |  |  |  | 0.850 | 0.031 | 1 | 1.607  | 0.041 | 1 |
| 94  | 17.131 | 0.019 | 1 | 1.430 | 0.020 | 1 |  |  |  | 0.770 | 0.016 | 1 |        |       |   |
| 95  | 16.637 | 0.001 | 1 | 2.267 | 0.009 | 1 |  |  |  | 1.315 | 0.005 | 2 | 2.531  | 0.003 | 1 |
| 96  | 18.460 | 0.032 | 1 | 1.583 | 0.035 | 1 |  |  |  | 0.933 | 0.027 | 1 |        |       |   |
| 97  | 18.422 | 0.015 | 1 | 1.884 | 0.024 | 1 |  |  |  | 1.02  | 0.014 | 1 | 1.874  | 0.016 | 1 |
| 98  | 18.950 | 0.038 | 1 | 1.597 | 0.044 | 1 |  |  |  | 1.061 | 0.032 | 1 | 1.872  | 0.041 | 1 |
| 99  | 18.867 | 0.039 | 1 | 1.744 | 0.044 | 1 |  |  |  | 0.987 | 0.033 | 1 |        |       |   |
| 100 | 18.557 | 0.006 | 1 | 2.249 | 0.029 | 1 |  |  |  |       |       |   |        |       |   |
| 101 | 19.075 | 0.008 | 1 | 1.777 | 0.029 | 1 |  |  |  | 1.038 | 0.008 | 1 | 1.577  | 0.016 | 1 |
| 102 | 19.189 | 0.012 | 1 | 1.697 | 0.033 | 1 |  |  |  | 1.037 | 0.012 | 1 | 1.955  | 0.016 | 1 |
| 103 | 19.205 | 0.009 | 1 | 1.727 | 0.033 | 1 |  |  |  | 1.024 | 0.009 | 1 | 2.142  | 0.014 | 1 |
| 104 | 19.546 | 0.011 | 1 | 1.611 | 0.038 | 1 |  |  |  | 1.121 | 0.012 | 1 | 2.004  | 0.02  | 1 |
| 105 | 19.709 | 0.013 | 1 | 1.782 | 0.049 | 1 |  |  |  | 1.001 | 0.012 | 1 | 1.6905 | 0.004 | 2 |

**Table 6.** (X, Y) position on the map of stars with photometry in ALICANTE 5. USNO-B1 identification for these stars and their coordinates.

| Number | RA (J2000) | DEC (J2000) | Name (USNO-B1.0) | X (Pixels) | Y (Pixels) |
|---|---|---|---|---|---|
| 1 | 09:16:19.78 | -49:44:30.1 | 0402-0143645 | 1339.87 | 784.48 |
| 2 | 09:16:25.29 | -49:41:52.6 | 0403-0142758 | 1990.52 | 2742.73 |
| 3 | 09:16:17.00 | -49:43:15.0 | 0402-0143604 | 1003.37 | 1711.26 |
| 4 | 09:16:25.72 | -49:45:11.7 | 0402-0143709 | 2059.72 | 274.40 |
| 5 | 09:16:15.43 | -49:45:28.6 | 0402-0143581 | 822.01 | 45.76 |
| 6 | 09:16:16.35 | -49:43:31.5 | 0402-0143598 | 920.69 | 1508.68 |
| 7 | 09:16:11.01 | -49:44:14.3 | 0402-0143536 | 283.70 | 970.04 |
| 8 | 09:16:15.54 | -49:43:01.7 | 0402-0143585 | 817.98 | 1866.87 |
| 9 | 09:16:10.16 | -49:45:19.9 | 0402-0143522 | 195.65 | 158.25 |
| 10 | 09:16:15.74 | -49:44:45.1 | 0402-0143593 | 849.07 | 604.87 |
| 11 | 09:16:10.18 | -49:43:42.4 | 0402-0143523 | 179.00 | 1371.38 |
| 12 | 09:16:13.11 | -49:41:31.7 | 0403-0142635 | 520.27 | 2991.12 |
| 13 | 09:16:24.57 | -49:42:11.8 | 0402-0143697 | 1904.67 | 2505.82 |
| 14 | 09:16:11.40 | -49:44:02.9 | 0402-0143537 | 327.79 | 1114.70 |
| 15 | 09:16:24.02 | -49:44:00.6 | 0402-0143692 | 1834.92 | 1165.90 |
| 16 | 09:16:22.01 | -49:44:23.8 | 0402-0143671 | 1607.89 | 867.26 |
| 17 | 09:16:09.83 | -49:43:29.0 | 0402-0143514 | 136.99 | 1532.80 |
| 18 | 09:16:13.61 | -49:45:05.4 | 0402-0143566 | 606.04 | 337.39 |
| 19 | 09:16:10.81 | -49:43:05.9 | 0402-0143533 | 255.17 | 1814.90 |
| 20 | 09:16:20.14 | -49:42:41.5 | 0402-0143648 | 1372.35 | 2134.66 |
| 21 | 09:16:12.23 | -49:44:58.2 | 0402-0143552 | 434.02 | 428.99 |
| 22 | 09:16:24.32 | -49:41:21.4 | 0403-0142750 | 1868.39 | 3134.39 |
| 23 | 09:16:24.46 | -49:43:20.1 | 0402-0143696 | 1896.46 | 1659.42 |
| 24 | 09:16:17.24 | -49:40:34.2 | 0403-0142696 | 1010.92 | 3709.96 |
| 25 | 09:16:10.93 | -49:41:02.4 | 0403-0142619 | 256.33 | 3351.77 |
| 26 | 09:16:09.66 | -49:41:07.1 | 0403-0142610 | 93.71 | 3294.47 |
| 27 | 09:16:15.46 | -49:41:53.2 | 0403-0142663 | 802.11 | 2728.31 |
| 28 | 09:16:24.12 | -49:44:36.1 | 0402-0143694 | 1862.80 | 713.55 |
| 29 | 09:16:23.53 | -49:40:17.7 | 0403-0142744 | 1765.50 | 3916.50 |
| 30 | 09:16:16.10 | -49:42:18.0 | 0402-0143595 | 879.16 | 2418.80 |
| 31 | 09:16:14.92 | -49:40:53.7 | 0403-0142654 | 736.25 | 3465.33 |
| 32 | 09:16:12.92 | -49:41:25.3 | 0403-0142633 | 493.83 | 3076.15 |
| 33 | 09:16:10.80 | -49:42:59.5 | 0402-0143532 | 254.25 | 1901.21 |
| 34 | 09:16:17.09 | -49:45:21.0 | 0402-0143606 | 1022.23 | 148.40 |
| 35 | 09:16:12.83 | -49:44:48.6 | 0402-0143554 | 508.00 | 549.17 |
| 36 | 09:16:16.11 | -49:40:02.9 | - | - | - |
| 37 | 09:16:25.75 | -49:42:44.7 | 0402-0143710 | 2052.43 | 2096.37 |
| 38 | 09:16:16.49 | -49:40:40.6 | 0403-0142681 | 921.30 | 3630.64 |
| 39 | 09:16:11.45 | -49:45:10.8 | 0402-0143538 | 342.80 | 271.60 |
| 40 | 09:16:23.55 | -49:40:16.2 | - | 1772.50 | 3937.50 |
| 41 | 09:16:24.00 | -49:42:35.4 | 0402-0143691 | 1838.80 | 2211.37 |
| 42 | 09:16:14.22 | -49:40:27.9 | 0403-0142644 | 648.05 | 3786.04 |
| 43 | 09:16:24.23 | -49:44:05.7 | 0402-0143695 | 1879.60 | 1083.05 |
| 44 | 09:16:14.27 | -49:44:35.7 | 0402-0143570 | 678.42 | 707.90 |
| 45 | 09:16:22.10 | -49:43:50.7 | 0402-0143674 | 1615.13 | 1275.32 |
| 46 | 09:16:21.18 | -49:45:22.0 | 0402-0143661 | 1515.47 | 141.38 |
| 47 | 09:16:29.71 | -49:42:49.6 | - | 2528.39 | 2043.51 |
| 48 | 09:16:29.78 | -49:42:45.7 | - | 2536.71 | 2091.25 |
| 49 | 09:16:38.10 | -49:43:34.3 | 0402-0143859 | 3541.29 | 1495.39 |
| 50 | 09:16:37.75 | -49:42:46.1 | 0402-0143855 | 3496.50 | 2096.50 |
| 51 | 09:16:27.67 | -49:43:02.5 | - | 2283.48 | 1880.85 |
| 52 | 09:16:34.24 | -49:43:19.3 | 0402-0143814 | 3077.78 | 1677.86 |
| 53 | 09:16:42.38 | -49:42:33.1 | 0402-0143918 | 4057.47 | 2259.70 |
| 54 | 09:16:30.42 | -49:42:43.5 | 0402-0143773 | 2620.32 | 2118.49 |
| 55 | 09:16:28.85 | -49:42:53.0 | 0402-0143747 | 2404.56 | 1996.56 |
| 56 | 09:16:28.85 | -49:42:53.0 | 0403-0142965 | 4236.50 | 2834.50 |
| 57 | 09:16:33.27 | -49:43:47.5 | 0402-0143804 | 2963.67 | 1327.31 |
| 58 | 09:16:30.54 | -49:44:42.6 | 0402-0143774 | 2638.59 | 639.61 |
| 59 | 09:16:34.48 | -49:40:55.6 | 0403-0142823 | 3087.09 | 3459.09 |
| 60 | 09:16:28.95 | -49:41:50.0 | 0403-0142784 | 2430.24 | 2783.87 |

**Table 6.** (*X*, *Y*) position on the map of stars with photometry in ALICANTE 5. USNO-B1 identification for these stars and their coordinates.

| Number | RA (J2000) | DEC (J2000) | Name (USNO-B1.0) | X (Pixels) | Y (Pixels) |
|---|---|---|---|---|---|
| 61 | 09:16:40.84 | -49:44:57.8 | 0402-0143895 | 3882.19 | 459.96 |
| 62 | 09:16:27.47 | -49:42:59.0 | 0402-0143733 | 2252.39 | 1938.32 |
| 63 | 09:16:36.38 | -49:41:44.9 | 0403-0142850 | 3326.43 | 2851.72 |
| 64 | 09:16:33.56 | -49:42:54.1 | 0402-0143807 | 2992.08 | 1988.50 |
| 65 | 09:16:33.14 | -49:43:10.3 | 0402-0143803 | 2944.15 | 1787.34 |
| 66 | 09:16:29.09 | -49:42:54.9 | - | 2453.96 | 1977.19 |
| 67 | 09:16:37.77 | -49:42:23.7 | 0402-0143856 | 3498.50 | 2368.50 |
| 68 | 09:16:38.10 | -49:43:34.3 | 0402-0143859 | 3192.90 | 1541.58 |
| 69 | 09:16:28.70 | -49:43:02.5 | 0402-0143745 | 2414.90 | 1872.54 |
| 70 | 09:16:39.35 | -49:42:49.5 | 0402-0143875 | 3691.43 | 2050.33 |
| 71 | 09:16:40.99 | -49:44:15.6 | 0402-0143896 | 3894.20 | 986.78 |
| 72 | 09:16:16.10 | -49:42:18.0 | 0402-0143595 | 828.96 | 2489.57 |
| 73 | 09:16:09.26 | -49:42:40.9 | 0402-0143501 | 60.38 | 2135.86 |
| 74 | 09:16:14.45 | -49:41:01.2 | 0403-0142646 | 680.81 | 3369.71 |
| 75 | 09:16:20.74 | -49:41:08.7 | 0403-0142721 | 1438.16 | 3286.42 |
| 76 | 09:16:20.87 | -49:40:34.9 | 0403-0142722 | 1451.87 | 3707.24 |
| 77 | 09:16:14.51 | -49:41:21.9 | 0403-0142648 | 687.31 | 3116.24 |
| 78 | 09:16:25.98 | -49:42:58.0 | 0402-0143719 | 2078.36 | 1934.04 |
| 79 | 09:16:11.30 | -49:40:55.0 | 0403-0142623 | 294.17 | 3452.13 |
| 80 | 09:16:13.07 | -49:43:07.9 | 0402-0143555 | 526.12 | 1798.89 |
| 81 | 09:16:19.98 | -49:42:29.6 | 0402-0143646 | 1352.35 | 2282.70 |
| 82 | 09:16:11.16 | -49:40:16.1 | 0403-0142620 | 279.15 | 3930.89 |
| 83 | 09:16:19.78 | -49:43:57.3 | 0402-0143644 | 1337.30 | 1192.50 |
| 84 | 09:16:13.55 | -49:42:23.1 | 0402-0143565 | 577.59 | 2355.85 |
| 85 | 09:16:14.97 | -49:40:20.5 | 0403-0142656 | 734.66 | 3873.97 |
| 86 | 09:16:18.60 | -49:45:31.9 | 0402-0143627 | 1204.33 | 15.83 |
| 87 | 09:16:22.75 | -49:43:41.7 | 0402-0143678 | 1695.50 | 1386.50 |
| 88 | 09:16:12.12 | -49:42:54.6 | 0402-0143550 | 409.89 | 1962.41 |
| 89 | 09:16:23.31 | -49:43:05.7 | 0402-0143683 | 1755.78 | 1837.96 |
| 90 | 09:16:13.91 | -49:42:03.3 | 0402-0143568 | 622.51 | 2598.34 |
| 91 | 09:16:09.44 | -49:45:30.0 | 0402-0143503 | 101.27 | 28.11 |
| 92 | 09:16:09.66 | -49:45:05.5 | 0402-0143508 | 129.47 | 331.69 |
| 93 | 09:16:37.81 | -49:40:18.7 | 0403-0142868 | 3486.26 | 3920.18 |
| 94 | 09:16:27.64 | -49:44:21.3 | 0402-0143737 | 2288.43 | 903.74 |
| 95 | 09:16:39.90 | -49:44:24.3 | 0402-0143881 | 3762.39 | 877.81 |
| 96 | 09:16:27.42 | -49:44:50.9 | - | 2261.74 | 537.13 |
| 97 | 09:16:42.99 | -49:41:38.0 | 0403-0142953 | 4122.86 | 2941.95 |
| 98 | 09:16:42.22 | -49:40:39.6 | 0403-0142943 | 4026.58 | 3668.36 |
| 99 | 09:16:27.69 | -49:45:05.5 | 0402-0143738 | 2296.00 | 353.13 |
| 100 | 09:16:30.38 | -49:43:05.3 | 0402-0143772 | 2610.79 | 1846.35 |
| 101 | 09:16:30.87 | -49:42:50.3 | 0402-0143782 | 2674.10 | 2031.16 |
| 102 | 09:16:43.65 | -49:42:17.7 | 0402-0143931 | 4206.81 | 2446.18 |
| 103 | 09:16:33.74 | -49:42:36.2 | 0402-0143808 | 3015.21 | 2217.83 |
| 104 | 09:16:41.70 | -49:42:59.3 | - | 3976.59 | 1932.24 |
| 105 | 09:16:30.64 | -49:42:07.8 | - | 2637.66 | 2562.09 |

**Table 8.** Photometry for membership of Pismis 11.

| $N^0$ | $V_0$ | $\sigma_{V_0}$ | $E(B-V)$ | $\sigma_{E(B-V)}$ | $(B-V)_0$ | $\sigma_{(B-V)_0}$ | Spectral Type |
|---|---|---|---|---|---|---|---|
| 2 | 11.601 | 0.037 | 1.234 | 0.018 | -0.190 | 0.059 | b7v |
| 5 | 10.700 | 0.028 | 1.212 | 0.017 | -0.227 | 0.093 | (b1.5-b2)v |
| 14 | 10.990 | 0.030 | 1.563 | 0.017 | -0.222 | 0.048 | b7v |
| 15 | 11.832 | 0.036 | 1.187 | 0.018 | -0.139 | 0.065 | b7v |
| 19 | 11.883 | 0.031 | 1.239 | 0.019 | -0.166 | 0.069 | b7v |
| 23 | 11.490 | 0.032 | 1.317 | 0.017 | -0.231 | 0.050 | b5v |
| 26 | 13.352 | 0.058 | 1.167 | 0.018 | -0.176 | 0.060 | b5v |
| 36 | 13.297 | 0.058 | 1.339 | 0.018 | -0.137 | 0.039 | b7v |
| 38 | 11.325 | 0.029 | 1.498 | 0.018 | -0.189 | 0.055 | b8v |
| 39 | 12.623 | 0.038 | 1.263 | 0.018 | -0.200 | 0.055 | b8v |
| 40 | 12.845 | 0.043 | 1.271 | 0.018 | -0.158 | 0.058 | b9v |
| 41 | 11.638 | 0.031 | 1.411 | 0.018 | -0.196 | 0.048 | b8v |
| 42 | 12.328 | 0.043 | 1.173 | 0.018 | -0.155 | 0.052 | b8v |
| 49 | 9.182 | 0.028 | 1.175 | 0.017 | -0.219 | 0.069 | b0.7v |
| 53 | 12.375 | 0.043 | 1.287 | 0.019 | -0.184 | 0.061 | b8v |
| 54 | 13.063 | 0.046 | 1.21 | 0.019 | -0.165 | 0.057 | b9v |
| 57 | 12.915 | 0.035 | 1.235 | 0.018 | -0.152 | 0.062 | b9v |
| 62 | 11.321 | 0.036 | 1.02 | 0.017 | -0.181 | 0.084 | b5v |
| 69 | 9.040 | 0.028 | 1.146 | 0.017 | -0.206 | 0.068 | b1v |
| 70 | 11.743 | 0.041 | 1.157 | 0.018 | -0.229 | 0.049 | b5v |
| 72 | 9.443 | 0.029 | 1.289 | 0.017 | -0.218 | 0.069 | b1.5v |
| 78 | 12.262 | 0.032 | 1.17 | 0.018 | -0.156 | 0.053 | b8v |
| 79 | 12.274 | 0.061 | 1.066 | 0.018 | -0.181 | 0.070 | b5v |
| 92 | 12.708 | 0.035 | 1.221 | 0.018 | -0.146 | 0.055 | b9v |
| 94 | 10.684 | 0.040 | 1.294 | 0.018 | -0.235 | 0.064 | b3v |
| 95 | 12.012 | 0.042 | 1.221 | 0.017 | -0.227 | 0.051 | b5v |
| 99 | 12.889 | 0.037 | 0.944 | 0.017 | -0.160 | 0.047 | b7v |
| 100 | 12.678 | 0.036 | 1.094 | 0.017 | -0.119 | 0.048 | b9v |
| 102 | 12.201 | 0.089 | 1.182 | 0.018 | -0.172 | 0.056 | b8v |
| 104 | 13.124 | 0.047 | 1.177 | 0.018 | -0.177 | 0.052 | b7v |
| 106 | 10.022 | 0.027 | 1.077 | 0.018 | -0.236 | 0.067 | b1.5v |
| 111 | 13.288 | 0.051 | 0.993 | 0.017 | -0.140 | 0.049 | b7v |
| 117 | 13.179 | 0.057 | 1.328 | 0.019 | -0.147 | 0.065 | b7v |
| 119 | 11.322 | 0.034 | 0.988 | 0.017 | -0.175 | 0.057 | b7v |
| 121 | 14.030 | 0.078 | 1.295 | 0.019 | -0.113 | 0.069 | b9v |
| 123 | 9.384 | 0.028 | 1.165 | 0.018 | -0.252 | 0.067 | b1v |
| 125 | 13.042 | 0.055 | 1.055 | 0.018 | -0.165 | 0.057 | b5v |
| 135 | 12.332 | 0.043 | 1.172 | 0.018 | -0.163 | 0.055 | b7v |
| 136 | 12.857 | 0.051 | 1.376 | 0.018 | -0.136 | 0.042 | b8v |
| 137 | 12.862 | 0.049 | 1.197 | 0.018 | -0.116 | 0.052 | b8v |
| 139 | 13.066 | 0.084 | 1.176 | 0.019 | -0.134 | 0.063 | b7v |
| 144 | 12.293 | 0.044 | 1.089 | 0.018 | -0.197 | 0.049 | b4v |
| 146 | 13.085 | 0.054 | 1.095 | 0.018 | -0.198 | 0.054 | b5v |

**Table 9.** Photometry for membership of Alicante 5.

| N$^0$ | E(B-V) | $\sigma_{E(B-V)}$ | A$_V$ | $\sigma_{A_V}$ | (B-V)$_0$ | $\sigma_{(B-V)_0}$ | V$_0$ | $\sigma_{V_0}$ | Spectral Type |
|---|---|---|---|---|---|---|---|---|---|
| 2 | 1.290 | 0.010 | 4.027 | 0.025 | -0.038 | 0.011 | 12.215 | 0.026 | b8v |
| 3 | 1.340 | 0.000 | 4.202 | 0.027 | -0.146 | 0.018 | 12.341 | 0.035 | b7v |
| 4 | 1.300 | 0.010 | 4.056 | 0.029 | -0.105 | 0.045 | 12.794 | 0.046 | a0v |
| 6 | 1.580 | 0.010 | 4.934 | 0.029 | -0.138 | 0.021 | 12.445 | 0.040 | b8v |
| 12 | 1.770 | 0.010 | 5.504 | 0.027 | -0.148 | 0.043 | 12.212 | 0.031 | b7v-b8v |
| 15 | 1.710 | 0.010 | 5.340 | 0.044 | -0.149 | 0.015 | 12.630 | 0.048 | b7v-b8v |
| 16 | 1.460 | 0.010 | 4.557 | 0.045 | -0.103 | 0.045 | 13.878 | 0.046 | b8v |
| 18 | 1.710 | 0.010 | 5.339 | 0.033 | -0.098 | 0.062 | 12.864 | 0.038 | b9v |
| 21 | 1.060 | 0.010 | 3.325 | 0.025 | -0.130 | 0.028 | 11.817 | 0.035 | b8v |
| 32 | 1.580 | 0.010 | 4.931 | 0.039 | -0.060 | 0.052 | 13.834 | 0.043 | b9v |
| 34 | 1.600 | 0.010 | 4.991 | 0.042 | -0.093 | 0.050 | 13.958 | 0.047 | b9v |
| 75 | 1.700 | 0.010 | 5.307 | 0.031 | -0.119 | 0.028 | 13.001 | 0.036 | a0v |
| 44 | 1.260 | 0.000 | 3.955 | 0.021 | -0.147 | 0.021 | 11.474 | 0.032 | b8v |
| 45 | 1.120 | 0.010 | 3.509 | 0.028 | -0.103 | 0.011 | 12.444 | 0.030 | b7v |
| 47 | 1.390 | 0.010 | 4.337 | 0.024 | -0.187 | 0.028 | 8.922 | 0.035 | b1v-b2v |
| 55 | 1.430 | 0.010 | 4.472 | 0.031 | -0.183 | 0.027 | 9.947 | 0.043 | b3v-b5v |
| 62 | 1.440 | 0.010 | 4.508 | 0.032 | -0.218 | 0.029 | 10.622 | 0.045 | b5v |
| 67 | 1.860 | 0.010 | 5.785 | 0.020 | -0.356 | 0.021 | 9.107 | 0.025 | b2v-b3v |
| 68 | 1.130 | 0.010 | 3.550 | 0.031 | -0.098 | 0.039 | 12.928 | 0.050 | b8v |
| 69 | 1.570 | 0.010 | 4.909 | 0.029 | -0.197 | 0.033 | 11.366 | 0.041 | b7v |
| 70 | 1.530 | 0.010 | 4.772 | 0.022 | -0.157 | 0.033 | 11.512 | 0.033 | b7v |
| 71 | 1.280 | 0.010 | 4.000 | 0.031 | -0.132 | 0.020 | 12.974 | 0.032 | b8v |